\documentclass[onecolumn,showpacs,preprintnumbers,amsmath,amssymb]{revtex4}
\usepackage{amsmath}
\usepackage{amssymb}
\usepackage{bm}
\usepackage{graphicx}
\usepackage{color}

\begin{document}

\title{ The Staggered Fermion for the Gross-Neveu Model at Non-zero Temperature and Density }

\author{Daming Li}
\email{lidaming@sjtu.edu.cn}

\affiliation{School of Mathematical Sciences, Shanghai Jiao Tong
University, Shanghai, 200240, China}

\date{\today}

\begin{abstract}
The 2+1d Gross-Neveu model with finite density and finite temperature are studied by the staggered fermion discretization. The kinetic part of this
staggered fermion in momentum space is used to build the relation between the staggered fermion and Wilson-like fermion. In the large $N_f$ limit (the number $N_f$ of staggered fermion flavors), the chiral condensate and fermion density are solved from the gap equation in momentum space, and thus the phase diagram of fermion coupling, temperature and chemical potential are obtained. Moreover, an analytic formula for the inverse of the staggered fermion matrix are given explicitly, which can be calculated easily by parallelization. The generalization to the 1+1d and 3+1d cases are also considered.
\end{abstract}



\maketitle

\section{Introduction}

The chiral phase transition in QCD from the hadronic phase at low temperature $T$ (low density $\mu_B$) to the quark-gluon plasma phase at high temperature (high density) has been studied intensively in the last decade. Although the relative firm statements for the phase structure can be made in two limit cases: finite $T$ with small baryon density $\mu_B \ll T$ and asymptotically high density $\mu_B \gg \Lambda_{\text{QCD}}$, the phase structures at the intermediate baryon density are not clear. For a recent review of QCD with finite density, see Ref. \cite{Fukushima_4814}\cite{Forcrand_0539}\cite{Schmidt_04707}.

Since the chiral symmetry breaking and restoration are intrinsically non-perturbative, the number of techniques are limited and most results comes from the lattice QCD. Unfortunately the lattice QCD at finite density suffers from the notorious sign problem, especially for the intermediate or large baryon density. For some simpler quantum field models, e.g.,
the dense two-color QCD \cite{Cotter_034507}, the sign problem can be avoided. The recent progress of the sign problem in lattice field models can refer to \cite{Daming_114501} and references therein.

This paper address a simplest four-fermion model with $Z_2$ symmetry: Gross-Neveu model at non-zero temperature and density
\cite{Rosenstein_59}\cite{Rosenstein_3088}\cite{Hands_9206024}\cite{Hands_9208022}\cite{Kogut_9904008}. The 2+1d Gross-Neveu model has an interesting continuum limit and there is a critical coupling indicating the threshold for the symmetry breaking at zero temperature and density.
Although the 2+1d Gross-Neveu model is not renormalisable in the weak coupling expansion, it is renormalisable
in $1/N_f$ expansion \cite{Rosenstein_59}, where $N_f$ is the number of flavors of fermions.

The symmetry breaking of Gross-Neveu model for the 1+1d case has been studied extensively
\cite{Wolff_303}\cite{Thies_2003}\cite{Forcrand_0610117}\cite{Castorina_2003}. Recently the domain wall fermion was used to study the symmetry breaking for 2+1d Gross-Neveu model at zero temperature and density \cite{Hands_1610.04394}.

Compared with the Wilson fermion, the staggered fermion are more adequate for studying spontaneous chiral symmetry breaking. Another advantage of the staggered fermion is due to the reduced computational cost since the Dirac matrices have been replaced by the staggered phase factor. The reconstruction of the Wilson-like fermion from the staggered fermion is rather technique and thus need a careful explanation of the physical fermions
for lattice QCD \cite{Rothe_2005} and for Gross-Neveu model \cite{Hands_9206024}.

In this paper we revisit the staggered fermion for the 1+1d, 2+1d and 3+1d Gross-Neveu model at non-zero temperature and finite density. The gap equation, which is based on the large $N_f$ limit, are solved in the momentum space. Moreover we derive an explicit formula for the inverse matrix
 of the staggered fermion matrix, which is easy to be implemented by parallelization and thus make the large scale calculation of the gap equation feasible.

The arrangement of the paper is as follows. The continuum 2+1d Gross-Neveu model at finite density and non-zero temperature is introduced in section \ref{The Gross-Neveu model}. In section \ref{The staggered fermion} the 2+1d staggered fermion is shown and non-dimensional quantities are introduced. The kinetic part of staggered fermion in the momentum space is given in section \ref{Staggered fermion in momentum space}, where the trace of the inverse matrix and elements of inverse matrix are given explicitly in momentum space. In section \ref{The 1+1d and 3+1d staggered fermion matrix}
the results in section \ref{Staggered fermion in momentum space} are generalized to the 1+1d and 3+1d staggered fermion matrix. Gap equation are given in section \ref{The gap equation}, where the chiral condensate and fermion density are calculated. The simulation results in the large $N_f$ limit
are obtained in section \ref{Simulation results}. Finally the conclusion are given in section \ref{Conclusion}.

\section{The Gross-Neveu model}\label{The Gross-Neveu model}

The Gross-Neveu (GN) model for interacting fermions in 2+1d is defined by the continuum Euclidiean Lagrangian density at finite density
\begin{eqnarray}\label{2016_12_4}
{\cal L}= \bar \psi ( \partial \!\!\!/ + \tilde \mu \gamma_0 +\tilde m)\psi - \frac{\tilde g^2}{2N_f}(\bar\psi\psi)^2
\end{eqnarray}
where $\partial \!\!\!/= \sum_{\nu=0}^2 \gamma_\nu \partial_\nu$, $\tilde \mu$ is the chemical potential, $\tilde m$ the bare mass, $\psi$ and $\bar \psi$ are an $N_f$-flavor 4 component spinor fields. Here we choose the Gamma matrices
\begin{eqnarray}\label{2017_9_12_20}
\gamma_\nu=\left( \begin{array}{cc}
\sigma_{\nu+1} & 0 \\
0 & -\sigma_{\nu+1}  \\
\end{array} \right), \quad \nu = 0,1,2
\end{eqnarray}
\begin{eqnarray}\label{2017_9_12_21}
\gamma_3 =\left( \begin{array}{cc}
 & - i\mathbb{I}_2 \\
 i\mathbb{I}_2 &   \\
\end{array} \right), \quad \gamma_5= \gamma_0\gamma_1\gamma_2\gamma_3 =  \left( \begin{array}{cc}
 &  \mathbb{I}_2 \\
 \mathbb{I}_2 &   \\
\end{array} \right)
\end{eqnarray}
where $\sigma_i(i=1,2,3)$ are the Pauli matrices. The Gamma matrices
satisfies
\begin{eqnarray*}
\gamma_\mu\gamma_\nu +  \gamma_\nu\gamma_\mu     = \delta_{\mu\nu} 2\mathbb{I}_4, \quad \mu,\nu = 0,1,2,3,5
\end{eqnarray*}
There is a discrete $Z_2$ symmetry $\psi\rightarrow \gamma_5\psi$,
$\bar\psi\rightarrow -\bar\psi\gamma_5$, which is broken by the mass term but not the interaction.
Introducing the bosonic filed $\sigma$, the interaction between fermions is decoupled,
\begin{eqnarray}\label{2016_12_4}
{\cal L}= \bar \psi ( \partial \!\!\!/ + \tilde \mu \gamma_0+ \tilde m+ \sigma)\psi + \frac{N_f}{2\tilde g^2}\sigma^2
\end{eqnarray}
The dimension of quantities for the 2+1d GN model is as follows
\begin{eqnarray}\label{2017_8_30_4}
 [\bar\psi] = [\psi] = [\tilde\mu] = [\tilde m] = [\sigma] = \text{length}^{-1} , \quad [\tilde g] = \text{length}^{1/2}
\end{eqnarray}
The partition function for this model is
\begin{eqnarray}\label{2017_8_30_1}
Z &=& \int d\bar\psi d\psi d \sigma e^{-\int\cal L} \nonumber  \\ &=&
\int d \sigma e^{-\int \frac{N_f}{2\tilde g^2}\sigma^2 } [\det(\partial \!\!\!/ +  \tilde\mu \gamma_0 + \tilde m+\sigma)]^{N_f} \nonumber \\
&=&
\int d \sigma \exp\Big(-\int \frac{N_f}{2\tilde g^2}\sigma^2 + N_f \ln [\det(\partial \!\!\!/+ \tilde\mu \gamma_0 +\tilde m+\sigma)]\Big)
\end{eqnarray}
where $\int \equiv \int_0^\beta dx_0 \int_0^L dx_1dx_2 $ with the inverse temperature $\beta=1/T$ and the space size $L$. $\bar\psi$ and $\psi$
are antiperiodic in $x_0$ direction, and are periodic in $x_1$ and $x_2$ directions.
We want to calculate the chiral condensate for one flavor
\begin{eqnarray}\label{2017_12_18_1}
\frac{1}{N_fV}\frac{\partial \ln Z}{\partial \tilde m}=
 \langle -\frac{1}{V}\int\bar\psi_i\psi_i \rangle   =
\frac{1}{\tilde g^2} \langle \frac{1}{V}\int\sigma \rangle \equiv \frac{1}{\tilde g^2} \Sigma
\end{eqnarray}
where $V=\beta L^2$ is the volume of 2+1d system. In the second equality we used
\begin{eqnarray*}
0 =\int d\bar\psi d\psi d \sigma \frac{\delta}{\delta\sigma(x)} e^{-\int\cal L} =
\int d\bar\psi d\psi d \sigma  e^{-\int\cal L} (-1)  (\bar\psi\psi + \frac{N}{\tilde g^2}\sigma )(x)
\end{eqnarray*}
Since the Lagrangian density is translation invariant, $\langle \bar\psi(x) \psi(x)\rangle$ and $\langle \sigma(x)\rangle$ does not
depend on $x$.
This model in the large $N_f$ limit can be solved exactly \cite{Hands_9206024} in the chiral limit $\tilde m=0$, which is based on the saddle approximation (gap equation) in
(\ref{2017_8_30_1})
\begin{eqnarray}\label{2017_8_30_9}
0& = & -\frac{V}{\tilde g^2}\Sigma + \frac{d}{d\Sigma} \ln [\det(\partial \!\!\!/+\tilde  \mu \gamma_0 + \tilde m+\Sigma)] \nonumber \\
&=& -\frac{V}{\tilde g^2}\Sigma   + \text{Tr} ( \partial \!\!\!/ + \tilde \mu \gamma_0+ \tilde m+\Sigma)^{-1} \nonumber \\
&=& -\frac{V}{\tilde g^2}\Sigma   + \sum_k\text{tr} ( i k \!\!\!/ + \tilde \mu \gamma_0+\tilde  m+\Sigma)^{-1} \nonumber \\
&=& -\frac{V}{\tilde g^2}\Sigma   + 4(\tilde m+\Sigma) \sum_k \Big( (k_0 -i \tilde \mu)^2+ \sum_{\nu=1,2}k_\nu^2+ (\tilde m+\Sigma)^2\Big)^{-1}
\end{eqnarray}
where in the third equality we write the trace of operator in momentum space and the summation over $k=(k_0,k_1,k_2)$
\begin{eqnarray*}
 k_0 = (2n-1)\pi T, \quad k_\nu = 2n_\nu \pi/L, \quad n, n_\nu \in \mathbf{Z}, \quad \nu=1,2
\end{eqnarray*}

\section{The staggered fermion}\label{The staggered fermion}
The staggered fermion discretization of the action $\int\cal L$ is
\begin{eqnarray}\label{2017_8_30_7}
S &=&  a^2 a_t \sum_{x,y}\bar\psi(x) \Big(
 \sum_{\alpha=1,2}
\frac{\eta_{x,\alpha}}{2a}(   \delta_{x+\hat\alpha,y}- \delta_{x,y+\hat\alpha})\Big) \psi(y)+ \nonumber\\
 && a^2 a_t \sum_{x,y}\bar\psi(x) \Big(
\frac{\eta_{x,0}}{2a_t}( e^{a_t\tilde \mu} s^1_x \delta_{x+\hat 0,y}-  e^{-a_t\tilde \mu} s^2_x \delta_{x,y+\hat 0})\Big) \psi(y)+ \nonumber\\ && a^2a_t \sum_x (\tilde m+\phi(x))\bar\psi(x)\psi(x)  + a^2 a_t \frac{N_f}{2\tilde g^2} \sum_{\tilde x}  \sigma(\tilde x)^2
\end{eqnarray}
with staggered phase factor $\eta_{x,0}=1$, $\eta_{x,1}=(-1)^{x_0/a}$,
$\eta_{x,2}=(-1)^{(x_0+x_1)/a}$.
$a N_x = L$, $a_tN_t = \beta = 1/T$.   The boundary condition for $\psi$ and
$\bar \psi$ are accounted for by the sign $s^1$ and $s^2$
\begin{eqnarray}
 s^1_{x} =  \left\{
  \begin{array}{l l}
-1  & \quad \text{if \ $x_0=N_t-1$}\\
1 &  \quad \text{Otherwise}\\
   \end{array} \right., \quad   s^2_{x} =  \left\{
  \begin{array}{l l}
-1  & \quad \text{if \  $x_0=0$}\\
1 &  \quad \text{Otherwise}\\
   \end{array} \right.
\end{eqnarray}
Here $\phi$ is defined on lattice $x$ by $\sigma(\tilde x)$
\begin{eqnarray}\label{2016_7_1_3}
 \phi(x) = \frac{1}{8} \sum_{[x,\tilde x]} \sigma(\tilde x) \quad
 \Longleftrightarrow  \quad \sigma(\tilde x) = \frac{1}{8} \sum_{[x,\tilde x]} \phi( x)
\end{eqnarray}
where $[x,\tilde x]$ denotes 8 dual lattices $\tilde x$ which is neighbour to $x$. The auxiliary field on dual lattice for two dimensional GN model was first studied in Ref. \cite{Cohen_102}.

According to (\ref{2017_8_30_4}), the non-dimensional quantities are introduced by
\begin{eqnarray}\label{2017_8_30_5}
 a  \sigma   \rightarrow \sigma, \quad
 a   \phi   \rightarrow \phi, \quad a\bar \psi  \rightarrow \bar\psi , \quad a  \psi  \rightarrow \psi
\end{eqnarray}
\begin{eqnarray}\label{2017_8_30_6}
  a  \tilde \mu  = \mu, \quad a \tilde m = m, \quad a^{-1/2}\tilde g  = g, \quad x/a\rightarrow x, \quad
 a_1 = a_t/a
\end{eqnarray}
and thus the action in (\ref{2017_8_30_7}) can be rewritten as
\begin{eqnarray*}
 S &=& a_1\sum_{x,y}\bar\psi(x) \Big(
 \sum_{\alpha=1,2}
\frac{\eta_{x,\alpha}}{2}(   \delta_{x+\hat\alpha,y}- \delta_{x,y+\hat\alpha})\Big) \psi(y)+ \\
 &&  \sum_{x,y}\bar\psi(x) \Big(
\frac{\eta_{x,0}}{2}(   e^{a_1\mu} s_x^1 \delta_{x+\hat 0,y}-  e^{-a_1\mu}s_x^2\delta_{x,y+\hat 0})\Big) \psi(y)+ \\ && a_1 \sum_x (m+\phi(x))\bar\psi(x)\psi(x)  + a_1 \frac{N_f}{2g^2} \sum_{\tilde x}  \sigma(\tilde x)^2
\end{eqnarray*}

The partition function for the Gross-Neveu model with $N_f$ flavors:
\begin{eqnarray}\label{2017_3_30_1}
Z= \int \prod_{i}   d\bar\psi_i d\psi_i d\sigma e^{-S}
\end{eqnarray}
where $\psi_i$ and $\bar\psi_i$ denote the Grassmann fields of
flavors $i=0,\cdots,N_f-1$ at the sites $x$, $\sigma$ is the real field define at the dual lattice sites $\tilde x$. The action is
\begin{eqnarray}\label{2017_3_12_1}
S=  \sum_{i,x,y}\bar\psi_i(x)D_{x,y}  \psi_i(y)  +
 \sum_{i,x} a_1\phi(x)\bar\psi_i(x)\psi_i(x)  +  \frac{a_1N_f}{2g^2} \sum_{\tilde x}  \sigma(\tilde x)^2
\end{eqnarray}
where
 \begin{eqnarray}\label{2017_7_31_1}
D_{x,y}&=& \left\{
  \begin{array}{l l}
a_1\frac{\eta_{x,\alpha}}{2}   & \quad \text{if \ $y=x+\hat\alpha$ }, \quad \alpha  = 1,2 \\
-a_1\frac{\eta_{x,\alpha}}{2}  & \quad \text{if \ $y=x-\hat\alpha$ }, \quad \alpha  = 1,2\\
\frac{\eta_{x,0}}{2}e^{a_1\mu
}s^1_{x}  & \quad \text{if \ $y=x+\hat 0$ } \\
-\frac{\eta_{x,0}}{2}e^{-a_1\mu
}s^2_{x}  & \quad \text{if \ $y=x-\hat 0$ }\\
a_1  m  & \quad \text{if \ $y=x$ }\\
0 &  \quad \text{otherwise}\\
   \end{array} \right.
\end{eqnarray}
The derivative of this matrix $D$ with respect to the chemical potential and bare mass are rather simple
 \begin{eqnarray*}
  \frac{\partial D_{x,y}}{\partial (a_1\mu)}     =   \frac{e^{a_1\mu} }{2}
 s^{1}_{x}\delta_{x+\hat 0,y}+\frac{e^{-a_1\mu} }{2}s^{2}_{x}\delta_{x,y+\hat 0}, \quad
  \frac{\partial D_{x,y}}{\partial (a_1m)}     =   \delta_{x,y}
\end{eqnarray*}
The real matrix $D(\mu,m)$ satisfies the following symmetry
$$D(\mu,m)_{x,y}=-D(-\mu,-m)_{y,x}  $$
$$ \varepsilon_x   D(\mu,m)_{x,y} \varepsilon_y  = -  D(\mu,-m)_{x,y}
=D(-\mu,m)_{y,x}
$$
where $ \varepsilon_x=(-1)^{x_0+x_1+x_2}$ is the parity of
site $x$.

By integrating the Grassmann fields, the partition function in (\ref{2017_3_30_1}) can be rewritten as
\begin{eqnarray}\label{2016_6_23_3}
Z = \int \prod_{\tilde x}d\sigma(\tilde x) e^{-S_{\textrm{eff}}}
\end{eqnarray}
with the effective action
\begin{eqnarray}\label{2016_6_23_4}
  S_{\textrm{eff}} = \frac{a_1N_f}{2g^2}\sum_{\tilde x}
\sigma^2(\tilde x)  -   N_f   \ln \det D[\phi]
\end{eqnarray}
and
\begin{eqnarray}\label{2016_7_31_3}
(D[\phi])_{x,y} = D_{x,y} + a_1 \phi(x) \delta_{x,y}
\end{eqnarray}
The computational results, e.g., non-dimensional chiral condensate and fermion density,  depend on the non-dimensional quantities
$$ (N_f, g, \mu, m, N_x,N_t) $$
The physical dimensional quantities can be recovered from the non-dimensional ones by introducing lattice size $a$ according to
(\ref{2017_8_30_5})(\ref{2017_8_30_6}). For notation simplicity, we set $a_t=a$ and thus $a_1=1$ in the following discussion.

\section{Staggered fermion in momentum space}\label{Staggered fermion in momentum space}
The kinetic part in (\ref{2017_3_12_1}) in one flavor is
$\sum_{x,y}\bar\chi(x)D_{x,y}\chi(y)$ where
\begin{eqnarray}\label{2017_8_30_11}
D_{x,y}&=& \left\{
  \begin{array}{l l}
\frac{\eta_{x,\alpha}}{2}   & \quad \text{if \ $y=x+\hat\alpha$ }, \quad \alpha  = 1,2 \\
-\frac{\eta_{x,\alpha}}{2}  & \quad \text{if \ $y=x-\hat\alpha$ }, \quad \alpha  = 1,2\\
\frac{\eta_{x,0}}{2}e^{\mu
}s^1_{x}  & \quad \text{if \ $y=x+\hat 0$ } \\
-\frac{\eta_{x,0}}{2}e^{-\mu
}s^2_{x}  & \quad \text{if \ $y=x-\hat 0$ }\\
 m  & \quad \text{if \ $y=x$ }\\
0 &  \quad \text{otherwise}\\
   \end{array} \right.
\end{eqnarray}
$\bar \chi$ and $\chi$ are the Grassmann fields defined on lattices. A Wilson-like fermion can be obtained from the
stagger fermion $\sum_{x,y}\bar\chi(x)D_{x,y}\chi(y)$ \cite{Hands_9206024}.

Assume that $N_x$ and $N_t$ are even integers. Let $Y=(Y_0,Y_1,Y_2)$ denotes a site on a lattice of twice the spacing of the original, and $A=(A_0,A_1,A_2), A_i=0,1$ is a lattice vector, which ranges over the corners of the elementary cube associated with $Y$, so that each site on the original lattice $x$ uniquely corresponds to $A$ and $Y$: $x=2Y+A$. Introducing notation
\begin{eqnarray*}
\chi(x) = \chi(2Y+A) = \chi(A,Y)
\end{eqnarray*}
A shift along $\mu$ direction can be represented by
\begin{eqnarray}\label{2017_9_13_3}
\chi(x+\hat\mu) & = & \chi(2Y+A+\hat\mu)= \chi(2(Y+\hat \mu)+A-\hat\mu) \nonumber \\ &=& \sum_{A^\prime}\Big(\delta_{A+\hat\mu,A^\prime}\chi(A^\prime,Y)
+ \delta_{A-\hat\mu,A^\prime}\chi(A^\prime,Y+\hat\mu)
 \Big)
\end{eqnarray}
Similarly,
\begin{eqnarray}\label{2017_9_13_2}
\chi(x-\hat\mu) = \sum_{A^\prime}\Big(\delta_{A-\hat\mu,A^\prime}\chi(A^\prime,Y)
+ \delta_{A+\hat\mu,A^\prime}\chi(A^\prime,Y-\hat\mu)
 \Big)
\end{eqnarray}
$\chi(x)$ is defined on the fine lattice sites $x$ with lattice size $a=1$
\begin{eqnarray}\label{2017_9_12_0}
 \{ x =   (x_0, x_1, x_2), \quad  0\leq x_0 < N_t, \ 0\leq x_1,x_2<N_x \}
\end{eqnarray}
 while $\chi(A,\cdot)$ on the coarse lattice sites $Y$ with lattice size $2a=2$
\begin{eqnarray}\label{2017_9_12_1}
 \{ 2Y =   2(Y_0, Y_1, Y_2), \quad  0\leq Y_0 < N_t/2, \ 0\leq Y_1,Y_2<N_x/2 \}
\end{eqnarray}
A unitary transformation of $\chi(A, \cdot)$ is defined by \cite{Burden_545}
\begin{eqnarray}\label{2017_8_30_20}
 u^{\alpha a}(Y)  = \frac{1}{4\sqrt{2}}\sum_A \Gamma_A^{\alpha a} \chi(A,Y), \quad
 d^{\alpha a}(Y)  = \frac{1}{4\sqrt{2}}\sum_A B_A^{\alpha a} \chi(A,Y)
\end{eqnarray}
\begin{eqnarray}\label{2017_8_30_21}
 \bar u^{\alpha a}(Y)  = \frac{1}{4\sqrt{2}}\sum_A \bar \chi(A,Y)\Gamma_A^{*\alpha a}  , \quad
 \bar  d^{\alpha a}(Y)  = \frac{1}{4\sqrt{2}}\sum_A \bar \chi(A,Y) B_A^{*\alpha a}
\end{eqnarray}
where $2\times 2$ matrices $\Gamma_A$ and $B_A$ is given by
\begin{eqnarray}\label{2017_9_12_5}
 \Gamma_A =  \sigma_1^{A_0}\sigma_2^{A_1}\sigma_3^{A_2}, \quad
B_A =  (-\sigma_1)^{A_0}(-\sigma_2)^{A_1}(-\sigma_3)^{A_2}
\end{eqnarray}
$\Gamma_A$ and $B_A$ satisfies the following properties (The indices $\alpha, \alpha^\prime, \beta, a, a^\prime$ and $b$ always run from 1 to 2)
\begin{eqnarray}\label{2017_8_30_22}
 \Gamma_{A\pm\hat\mu} = \eta_\mu(A)\sigma_{\mu+1} \Gamma_A , \quad  B_{A\pm\hat\mu} = \eta_\mu(A)(-\sigma_{\mu+1}) B_A, \quad \mu = 0,1,2
\end{eqnarray}
\begin{eqnarray}\label{2017_9_13_0}
\text{Tr}(\Gamma_A^\dagger \Gamma_{A^\prime} + B_A^\dagger B_{A^\prime}) = 4 \delta_{A{A^\prime}}
\end{eqnarray}
\begin{eqnarray}\label{2017_8_30_23}
 \sum_A\Gamma_{A}^{\alpha a}\Gamma_{A}^{*\beta b} =
\sum_AB_{A}^{\alpha a}B_{A}^{*\beta b}= 4\delta_{\alpha\beta}\delta_{ab}, \quad
\sum_A\Gamma_{A}^{\alpha a}B_{A}^{*\beta b} = \sum_AB_{A}^{\alpha a}\Gamma_{A}^{*\beta b} =0
\end{eqnarray}
\begin{eqnarray}\label{2017_8_30_24}\sum_{A,A_\mu=1}\Gamma_{A}^{\alpha a } (\Gamma_{A}^*)^{\alpha^\prime a^\prime} =
\sum_{A,A_\mu=0}\Gamma_{A}^{\alpha a } (\Gamma_{A}^*)^{\alpha^\prime a^\prime}, \quad \mu=0,1,2
\end{eqnarray}
Eq. (\ref{2017_8_30_24}) is also valid if $\Gamma$ is replaced by $B$.
\begin{eqnarray}\label{2017_9_6_1}
\sum_{A,A_\mu=1}\Gamma_{A}^{\alpha a } (\sigma_{\mu+1}^*B_{A}^*)^{\alpha^\prime a^\prime}
= - 2 \sigma_{\mu+1}^{*aa^\prime}\delta_{\alpha\alpha^\prime}
\end{eqnarray}
\begin{eqnarray}\label{2017_9_6_1_1}
\sum_{A,A_\mu=0}\Gamma_{A}^{\alpha a } (\sigma_{\mu+1}^*B_{A}^*)^{\alpha^\prime a^\prime}
=  2 \sigma_{\mu+1}^{*aa^\prime}\delta_{\alpha\alpha^\prime}
\end{eqnarray}
See \ref{Appendix_1} for these properties.

 Using (\ref{2017_9_13_0}),
the inverse transformation of (\ref{2017_8_30_20}) and (\ref{2017_8_30_21}) are
\begin{eqnarray}\label{2017_8_30_22}
\chi(A,Y) &=& \sqrt{2}\sum_{\alpha,a}\Big[\Gamma_A^{*\alpha a} u^{\alpha a}(Y) + B_A^{*\alpha a} d^{\alpha a}(Y)\Big]\\
\bar\chi(A,Y) &=& \sqrt{2}\sum_{\alpha,a}\Big[\bar u^{\alpha a}(Y)\Gamma_A^{\alpha a} + \bar d^{\alpha a}(Y)B_A^{\alpha a}\Big]
\end{eqnarray}
Let us introduce the two Dirac fields with 4 components ($a=1,2$)
\begin{eqnarray*}
q^a(Y) = \left( \begin{array}{c}
q_1^{ a}(Y) \\
q_2^{ a}(Y) \\
\end{array} \right)= \left( \begin{array}{c}
u^{\alpha a}(Y) \\
d^{\alpha a}(Y) \\
\end{array} \right), \quad
\bar q^a(Y) =  (\bar q_1^{ a}(Y), \bar q_2^{a}(Y))= (\bar u^{\alpha a}(Y), \bar d^{\alpha a}(Y))
\end{eqnarray*}
From the properties (\ref{2017_8_30_23}), it is easy to show that
\begin{eqnarray*}
&& \sum_{x}\bar\chi(x)\chi(x)\\ &=& \sum_{A,Y}\sqrt{2}\sum_{\alpha,a}\Big(\bar u^{\alpha a}(Y)\Gamma_A^{\alpha a} + \bar d^{\alpha a}(Y)B_A^{\alpha a}\Big) \sqrt{2}\sum_{\alpha^\prime,a^\prime}\Big[\Gamma_A^{*\alpha^\prime a^\prime} u^{\alpha^\prime a^\prime}(Y) + B_A^{*\alpha^\prime a^\prime} d^{\alpha^\prime a^\prime}(Y)\Big] \\
&=& 8\sum_{Y} \sum_{\alpha,a} \Big(  \bar u^{\alpha a}(Y)u^{\alpha a}(Y) + \bar d^{\alpha a}(Y)d^{\alpha a}(Y) \Big)\\
&= & 8  \sum_{Y,a} \bar q^a(Y) q^a(Y)= 8\sum_{Y}  \bar q(Y) q(Y)  = 8\sum_k \bar q(k) q(k)
 \end{eqnarray*}
where in the last equality the inner produce between $\bar q$ and $q$ is given in momentum space corresponding to the coarse lattice with
lattice size 2
\begin{eqnarray}\label{2017_8_30_25}
k = 2\pi \Big(\frac{m_0+\frac{1}{2}}{N_t},\frac{m_1}{N_x},\frac{m_2}{N_x}\Big), \quad 0\leq m_0 < N_t/2, \ 0\leq m_1,m_2<N_x/2
\end{eqnarray}

For any fixed $\mu=0,1,2$,
\begin{eqnarray*}
&& \frac{1}{2}
\sum_{x}\eta_\mu(x)\bar\chi(x)(\chi(x+\hat\mu)- \chi(x-\hat\mu)) \\ & =  &
\frac{1}{2}
\sum_{A,A^\prime,Y}\eta_\mu(A)\bar \chi(A,Y)\Big( \delta_{A+\hat\mu,A^\prime} (\chi(A^\prime,Y)-\chi(A^\prime,Y-\hat\mu))
+  \delta_{A-\hat\mu,A^\prime} (\chi(A^\prime,Y+\hat\mu)-\chi(A^\prime,Y)) \Big)  \\ & =  &
\frac{1}{2}
\sum_{A,A^\prime,Y}\eta_\mu(A)\bar \chi(A,Y)\Big( \frac{\delta_{A+\hat\mu,A^\prime} + \delta_{A-\hat\mu,A^\prime}}{2} \partial_\mu \chi(A^\prime,Y) +  \frac{\delta_{A-\hat\mu,A^\prime} - \delta_{A+\hat\mu,A^\prime}}{2} \partial_\mu^2 \chi(A^\prime,Y) \Big)  \\
   & =  & \frac{1}{2}\sum_{A,A^\prime,Y}\eta_\mu(A)
  \sqrt{2}\sum_{\alpha,a}\Big(\bar u^{\alpha a}(Y)\Gamma_A^{\alpha a} + \bar d^{\alpha a}(Y)B_A^{\alpha a}\Big) \\
   &&\Big\{
   \frac{\delta_{A+\hat\mu,A^\prime} + \delta_{A-\hat\mu,A^\prime}}{2}   \sqrt{2}\sum_{\alpha^\prime,a^\prime}\Big(\Gamma_{A^\prime}^{*\alpha^\prime a^\prime} \partial_\mu u^{\alpha^\prime a^\prime}(Y) + B_{A^\prime}^{*\alpha^\prime a^\prime} \partial_\mu  d^{\alpha^\prime a^\prime}(Y)\Big) + \\
 & & \frac{\delta_{A-\hat\mu,A^\prime} - \delta_{A+\hat\mu,A^\prime}}{2}  \sqrt{2}\sum_{\alpha^\prime,a^\prime}\Big(\Gamma_{A^\prime}^{*\alpha^\prime a^\prime}
\partial_\mu^2  u^{\alpha^\prime a^\prime}(Y)  +
  B_{A^\prime}^{*\alpha^\prime a^\prime}
  \partial_\mu^2  d^{\alpha^\prime a^\prime}(Y)\Big)\Big\}
\end{eqnarray*}
where in the second equality (\ref{2017_9_13_3}) and (\ref{2017_9_13_2}) are used.
According to the properties of $\Gamma_A$ and $B_A$ in (\ref{2017_8_30_23})
(\ref{2017_8_30_24})(\ref{2017_9_6_1}) and (\ref{2017_9_6_1_1})
\begin{eqnarray}\label{2017_10_9_1}
&& \frac{1}{2}
\sum_{x}\eta_\mu(x)\bar\chi(x)(\chi(x+\hat\mu)- \chi(x-\hat\mu)) \nonumber \\ & =  &
 2\sum_{Y}\Big(\bar u^{\alpha a}(Y)(\sigma_{\mu+1})^{\alpha\alpha^\prime } \delta_{a a^\prime}\partial_\mu u^{\alpha^\prime a^\prime}(Y)
 +    \bar d^{\alpha a}(Y)(-\sigma_{\mu+1})^{\alpha\alpha^\prime } \delta_{a a^\prime}\partial_\mu  d^{\alpha^\prime a^\prime}(Y)+ \nonumber  \\ &&
 \bar u^{\alpha a}(Y)(\sigma_{\mu+1}^*)^{aa^\prime } \delta_{\alpha \alpha^\prime}
\partial_\mu^2  d^{\alpha^\prime a^\prime}(Y) +  \bar d^{\alpha a}(Y)(-\sigma_{\mu+1}^*)^{aa^\prime } \delta_{\alpha \alpha^\prime}\partial_\mu^2  u^{\alpha^\prime a^\prime}(Y) \Big) \nonumber \\
& = & 2\sum_{Y} \Big[ \bar q(Y) (\gamma_\mu\otimes \mathbb{I}_2)\partial_\mu q(Y) +
\bar q(Y)   ( i\gamma_3 \otimes \sigma_{\mu+1}^*) \partial_\mu^2 q(Y) \Big ] \nonumber \\
& = & 8\sum_{Y} \Big[ \bar q(Y) (\gamma_\mu\otimes \mathbb{I}_2)\frac{\partial_\mu q(Y)}{4} +
\bar q(Y)   ( i\gamma_3 \otimes \sigma_{\mu+1}^*)\frac{\partial_\mu^2 q(Y)}{4} \Big ]\nonumber \\
& = & 8 \sum_k \Big[ \bar q(k) (\gamma_\mu\otimes \mathbb{I}_2)\frac{i}{ 2}\sin (2k_\mu) q(k) +
\bar q(k)   (i\gamma_3 \otimes \sigma_{\mu+1}^*) \frac{1}{ 2} [\cos(2k_\mu)-1] q(k) \Big ]
\end{eqnarray} where we used the notations
$$  \partial_\mu q(Y) = q(Y+\hat \mu) -  q(Y-\hat \mu) $$
$$  \partial^2_\mu q(Y) = q(Y+\hat \mu) - 2 q(Y) +  q(Y-\hat \mu) $$
and the summation over $k$ is taken for all modes in (\ref{2017_8_30_25}).
Similarly, we have (\ref{Appendix_0})
\begin{eqnarray}\label{2017_9_2_1}
& &\frac{1}{2}
\sum_{x}\bar\chi(x)(\chi(x+\hat 0)+ \chi(x-\hat 0)) \nonumber \\
& = & 8 \sum_k \Big[ \bar q(k)  (  i\gamma_3 \otimes \sigma_1^* ) i 2^{-1}\sin (2k_0) q(k) +
\bar q(k)    ( \gamma_0 \otimes \mathbb{I}_2 )  2^{-1} [\cos(2k_0)+1] q(k) \Big ]  \nonumber \\ & \equiv &8 \sum_k \bar q(k) A_+(k) q(k)
\end{eqnarray}

Using
\begin{eqnarray*}
&&\frac{1}{2}
\sum_{x}\bar\chi(x)(e^{\mu}\chi(x+\hat 0)- e^{-\mu}\chi(x-\hat 0)) \\ &=& \cosh\mu \Big[\frac{1}{2}
\sum_{x}\bar\chi(x)(\chi(x+\hat 0)- \chi(x-\hat 0))\Big] +  \\ && \sinh\mu \Big[\frac{1}{2}
\sum_{x}\bar\chi(x)(\chi(x+\hat 0)+ \chi(x-\hat 0))\Big]
\end{eqnarray*}
and (\ref{2017_10_9_1})(\ref{2017_9_2_1}), the kinetic part $\sum_{x,y}\bar\chi(x)D_{x,y}\chi(y)$ can be rewritten as in the momentum space
\begin{eqnarray}\label{2017_8_30_26}
\sum_{x,y}\bar\chi(x)D_{x,y}\chi(y) =  8\sum_k  \bar q(k) D(k) q(k)
\end{eqnarray}
where the summation over $k$ is taken for all momentum mode of coarse lattice according to (\ref{2017_8_30_25}), and the staggered matrix
in the momentum space is diagonal
\begin{eqnarray}\label{2017_8_30_27}
  D(k) &=& m + \sum_{\mu=1,2}\frac{i}{2}  \Big\{ (\gamma_\mu\otimes \mathbb{I}_2)\sin (2k_\mu)  +  (  \gamma_3 \otimes \sigma_{\mu+1}^*)  [\cos(2k_\mu)-1] \Big\}+ \nonumber\\
  &  & \frac{1}{2}\Big\{(\gamma_0\otimes \mathbb{I}_2)\Big(i\cosh\mu\sin(2k_0) + \sinh\mu [\cos(2k_0)+1]\Big) +\nonumber\\
  &  & (\gamma_3\otimes \sigma_1^*)\Big(i \cosh\mu[\cos(2k_0)-1] -   \sinh\mu \sin(2k_0)\Big)\Big\} \nonumber \\
  & \equiv & m + \sum_{\mu=0,1,2} (\gamma_\mu\otimes \mathbb{I}_2) a_\mu + \sum_{c=1,2,3} (\gamma_3\otimes \sigma_{c}^*) b_c
\end{eqnarray}
where $a_\mu$ and $b_c$ depends on $k$.
The inverse matrix of $D(k)$ is
\begin{eqnarray}\label{2017_8_30_29}
  D(k)^{-1} = \frac{1}{N(k)} \Big[ m - \sum_{\mu=0,1,2} (\gamma_\mu\otimes \mathbb{I}_2) a_\mu - \sum_{c=1,2,3} (\gamma_3\otimes \sigma_{c}^*) b_c
  \Big]
\end{eqnarray}
where
\begin{eqnarray}\label{2017_8_30_30}
N(k) & = &  m^2 + \frac{1}{4}\sum_{\mu=0,1,2} (\sin 2k_\mu)^2  + \frac{1}{4}\sum_{\mu=0,1,2} (1-\cos 2k_\mu)^2 \nonumber  \\
&&  -  \sinh^2\mu \cos 2k_0  -  i \cosh\mu\sinh\mu \sin 2k_0
 \end{eqnarray}

We can calculate the trace of inverse matrix $D$ in (\ref{2017_8_30_11}) from (\ref{2017_8_30_26})
\begin{eqnarray}\label{2017_8_30_31}
\sum_x D^{-1}_{x,x} & = & - \frac{\int e^{-\bar\chi D\chi} \sum_x \bar\chi (x) \chi(x) }{\int e^{-\bar\chi D\chi}} \nonumber  \\
& = & - \frac{\int e^{-\sum_k \bar q(k) 8 D(k) q(k)} 8 \sum_k  \bar q (k) q(k) }{\int  e^{-\sum_k \bar q(k) 8D(k) q(k)}}\nonumber  \\
& = & -8 \sum_k  \frac{\int e^{-  \bar q(k) 8D(k) q(k)}  \bar q(k)  q(k) }{\int  e^{-  \bar q(k) 8D(k) q(k)}} \nonumber  \\
 & = & 8\sum_k  \text{tr} [(8 D(k))^{-1}]  =  \sum_k \text{tr} [ D(k)^{-1}]= \sum_k \frac{8m}{N(k)}
\end{eqnarray}
where the summation over $k$ is given by (\ref{2017_8_30_25}). Note that the right hand side of (\ref{2017_8_30_31})
is real since $\sum_{k_0} \sin 2k_0/|N(k)|^2 =  0$ for any $k_1$ and $k_2$ modes in (\ref{2017_8_30_25}).
Similarly,
\begin{eqnarray}\label{2017_9_1_4_1}
 \sum_x (D^{-1}_{x+\hat 0,x}s^1_x + D^{-1}_{x-\hat 0,x}s^2_x)   = 8\sum_k \frac{b_1  \sin 2k_0 -  a_0 (\cos 2k_0 +1)}{N(k)}
\end{eqnarray}
and
\begin{eqnarray}\label{2017_9_1_10}
 \sum_x (D^{-1}_{x+\hat 0,x}s^1_x - D^{-1}_{x-\hat 0,x}s^2_x)  =  (-8 i) \sum_k \frac{a_0   \sin 2k_0 +  b_1 (\cos 2k_0-1 )}{N(k)}
\end{eqnarray}
The inverse matrix of $D$ in (\ref{2017_8_30_11}) is
\begin{eqnarray}\label{2017_9_13_10}
  D^{-1}_{x^\prime,x}
& = & \frac{1}{4} \sum_{\alpha,a,\alpha^\prime,a^\prime} \Big[\Gamma_A^{\alpha a}\Gamma_{A^\prime}^{*\alpha^\prime a^\prime}
D^{-1}_{(Y^\prime a^\prime \alpha^\prime 1; Y a  \alpha 1)} +
 \Gamma_A^{\alpha a}B_{A^\prime}^{*\alpha^\prime a^\prime} D^{-1}_{(Y^\prime a^\prime \alpha^\prime 2; Y a  \alpha 1)} + \nonumber \\
&&\hspace{1.2cm} B_A^{\alpha a}\Gamma_{A^\prime}^{*\alpha^\prime a^\prime}
D^{-1}_{(Y^\prime a^\prime \alpha^\prime 1; Y a  \alpha 2)} +  B_A^{\alpha a}B_{A^\prime}^{*\alpha^\prime a^\prime}
D^{-1}_{(Y^\prime a^\prime \alpha^\prime 2; Y a  \alpha 2)} \Big]
\end{eqnarray}
See \ref{Appendix_0_1} for the derivation of (\ref{2017_9_1_4_1})(\ref{2017_9_1_10})(\ref{2017_9_13_10}).

Since $D$ is diagonal in momentum space, the inverse matrix in the $\bar qq$ basis is
\begin{eqnarray*}
&& D^{-1}_{Y^\prime;Y} \\
 &=& \frac{1}{N_t/2(N_x/2)^2} \sum_k e^{i k \cdot 2(Y^\prime - Y)} D^{-1}(k)  \\
 & = &\frac{1}{N_t/2(N_x/2)^2} \sum_k e^{i k \cdot 2(Y^\prime - Y)} \frac{1}{N(k)} \Big[ m - \sum_{\mu=0,1,2} (\gamma_\mu\otimes \mathbb{I}_2) a_\mu - \sum_{c=1,2,3} (\gamma_3\otimes \sigma_{c}^*) b_c
  \Big]  \\
 & \equiv & m (\mathbb{I}_4\otimes \mathbb{I}_2) \tilde 1  (Y^\prime - Y)   -   \sum_{\mu=0,1,2}(\gamma_\mu\otimes \mathbb{I}_2) \tilde a_\mu (Y^\prime - Y) - \sum_{c=1,2,3} (\gamma_3\otimes \sigma_{c}^*)\tilde  b_c (Y^\prime - Y)
 \end{eqnarray*}
where the notation with tilde denotes the inverse Fourier transformation, e.g.,
\begin{eqnarray*}
\tilde a_\mu (Y) &=&  \frac{1}{\frac{N_t}{2}\Big(\frac{N_x}{2}\Big)^2}  \sum_k e^{i k \cdot 2Y}\frac{a_\mu(k)}{N(k)}  \nonumber \\
 & = & e^{i \frac{2\pi Y_0}{N_t}} \frac{1}{\frac{N_t}{2}\Big(\frac{N_x}{2}\Big)^2}  \sum_{m_0=0}^{\frac{N_t}{2}-1} \sum_{m_1=0}^{\frac{N_x}{2}-1}\sum_{m_2=0}^{\frac{N_x}{2}-1}
 e^{i2\pi \Big(\frac{m_0Y_0}{N_t/2}+\frac{m_1Y_1}{N_x/2}+\frac{m_2Y_2}{N_x/2}\Big)} \frac{a_\mu(m_0,m_1,m_2)}{N(m_0,m_1,m_2)}
\end{eqnarray*}
for $|Y_0| \leq \frac{N_t}{2}-1$, $|Y_1|,|Y_2| \leq \frac{N_x}{2}-1$. We first use the fast Fourier transformation to calculate $\tilde a_\mu (Y)\exp(-i \frac{2\pi Y_0}{N_t})$ and thus $\tilde a_\mu (Y)$ for $0\leq Y_0\leq\frac{N_t}{2}-1$, $0\leq Y_1,Y_2\leq \frac{N_x}{2}-1$. Then
$\tilde a_\mu (Y)$ for $|Y_0| \leq \frac{N_t}{2}-1$, $|Y_1|,|Y_2| \leq \frac{N_x}{2}-1$ can be obtained since it is anti-periodic in $Y_0$ direction and periodic in $Y_1$ and $Y_2$ direction.

Each term in $D^{-1}_{Y^\prime;Y}$ has a tensor product $A\otimes B$ between $4\times 4$ matrix $A= (A_{ij})_{i,j=1,2} $ with $2\times 2$ matrix $A_{ij}$ and $2\times 2$ matrix $B$. The indices of $D^{-1}_{(Y^\prime a^\prime \alpha^\prime i; Y a  \alpha j)}$ of the inverse matrix $D^{-1}_{Y^\prime;Y}$ in (\ref{2017_9_13_10}) is related to $(A_{ij})_{\alpha^\prime \alpha}B_{a^\prime a}$. The analytic formula for the
inverse matrix of the staggered fermion is the main contribution of this paper. Compared to the computational complexity $O((N_tN_x^2)^3)$ of the usual inverse matrix, the computational cost is $O(16 (N_tN_x^2)^2)$ since each element of the inverse matrix needs the summation over
$\alpha,a,\alpha^\prime,a^\prime=1,2$. Moreover a parallel implementation can be realized easily for the formula (\ref{2017_9_13_10}).

The trace of the inverse matrix in (\ref{2017_8_30_31}) can be derived from (\ref{2017_9_13_10})
\begin{eqnarray*}
 \sum_x D^{-1}_{x,x} =  \sum_{\alpha,a} \Big[D^{-1}_{(Y a \alpha 1; Y a  \alpha 1)} +
D^{-1}_{(Y a \alpha 2; Y a  \alpha 2)} \Big] = \sum_k \frac{8m}{N(k)}
\end{eqnarray*}

 \section{The 1+1d and 3+1d staggered fermion matrix}\label{The 1+1d and 3+1d staggered fermion matrix}
The staggered fermion matrix in (\ref{2017_8_30_11}) can be generalized to the 1+1d and 3+1d case, where $\alpha$ is 1 for the 1+1d case and
$\alpha$ run from 1 to 3 for the 3+1d case.

 For the 1+1d case, the $2\times 2$ matrices $\gamma_\mu$ are defined to be
\begin{eqnarray*}
\gamma_\mu = \sigma_\mu, \quad \mu = 1,2,  \quad \gamma_5 = i \gamma_1\gamma_2,  \quad
\gamma_\mu\gamma_\nu +  \gamma_\nu\gamma_\mu     = \delta_{\mu\nu} 2\mathbb{I}_2, \quad \mu,\nu = 1,2,5
\end{eqnarray*}
The unitary transformation in (\ref{2017_8_30_20}) and (\ref{2017_8_30_21}) are modified to be
\begin{eqnarray*}
 \psi^{\alpha a}(Y)  = \frac{1}{2}\sum_A \Gamma_A^{\alpha a} \chi(A,Y), \quad
 \bar \psi^{\alpha a}(Y)  = \frac{1}{2}\sum_A \bar \chi(A,Y)\Gamma_A^{*\alpha a}
\end{eqnarray*}
The kinetic part $\sum_{x,y}\bar\chi(x)D_{x,y}\chi(y)$ can be written as
\begin{eqnarray}\label{2017_9_25_4}
\sum_{x,y}\bar\chi(x)D_{x,y}\chi(y) = \sum_k  \bar \psi(k) D(k) \psi(k)
\end{eqnarray}
where the summation is taken over all modes
\begin{eqnarray}\label{2017_9_25_3}
k = 2\pi \Big(\frac{m_0+\frac{1}{2}}{N_t},\frac{m_1}{N_x}\Big), \quad 0\leq m_0 < N_t/2, \ 0\leq m_1<N_x/2
\end{eqnarray}
The fermion matrix in momentum space is diagonal
\begin{eqnarray}\label{2017_9_25_6}
  D(k) &=& 2m + \sum_{\mu=1} \Big\{ (\gamma_{\mu+1}\otimes \mathbb{I}_4)i\sin (2k_\mu)  +  (  \gamma_5 \otimes \gamma_{\mu+1}^*\gamma_5^*)  [\cos(2k_\mu)-1] \Big\}+ \nonumber\\
  &  & \Big\{(\gamma_1\otimes \mathbb{I}_4)\Big(i\cosh\mu\sin(2k_0) + \sinh\mu [\cos(2k_0)+1]\Big) +\nonumber\\
  &  & (\gamma_5\otimes \gamma_1^*\gamma_5^*)\Big( \cosh\mu[\cos(2k_0)-1] + i  \sinh\mu \sin(2k_0)\Big)\Big\} \nonumber \\
  & \equiv & 2m + \sum_{\mu=0,1} (\gamma_{\mu+1}\otimes \mathbb{I}_4) a_\mu + \sum_{\mu=0,1} (\gamma_5\otimes \gamma_{\mu+1}^*\gamma_5^*) b_\mu
\end{eqnarray}
with its inverse
\begin{eqnarray}\label{2017_9_25_7}
  D(k)^{-1} = \frac{1}{N(k)} \Big[ 2m - \sum_{\mu=0,1} (\gamma_{\mu+1}\otimes \mathbb{I}_4) a_\mu - \sum_{\mu=0,1} (\gamma_5\otimes \gamma_{\mu+1}^*\gamma_5^*) b_\mu   \Big]
\end{eqnarray}
where
\begin{eqnarray}\label{2017_9_25_9}
N(k) &=&  4m^2 + \sum_{\mu=1} (\sin 2k_\mu)^2 - \Big(i\cosh\mu\sin 2k_0 + \sinh\mu(\cos 2k_0+1)\Big)^2 \nonumber  \\
&& + \sum_{\mu=1} (1-\cos 2k_\mu)^2 +\Big(\cosh\mu(\cos 2k_0-1) + i\sinh\mu\sin 2k_0\Big)^2
\end{eqnarray}
The trace of the inverse matrix is
\begin{eqnarray}\label{2017_9_25_10}
\sum_x D^{-1}_{x,x}  = \sum_k \frac{16m}{N(k)}
\end{eqnarray}
The inverse matrix of $D$ in can be calculated
\begin{eqnarray}\label{2017_9_25_11}
  D^{-1}_{x^\prime,x} =  \sum_{\alpha,a,\alpha^\prime,a^\prime}\Gamma_A^{\alpha a}\Gamma_{A^\prime}^{*\alpha^\prime a^\prime}
D^{-1}_{(Y^\prime a^\prime \alpha^\prime; Y a  \alpha)}
\end{eqnarray}
where
\begin{eqnarray}\label{2017_9_25_12}
D^{-1}_{Y^\prime;Y} = \frac{1}{N_t/2(N_x/2)} \sum_k e^{i k \cdot 2(Y^\prime - Y)} D^{-1}(k)
\end{eqnarray}

 For the 3+1d case, the $4\times 4$ matrices $\gamma_\mu$ are defined to be
\begin{eqnarray*}
 \Gamma_A =  \gamma_1^{A_0}\gamma_2^{A_1}\gamma_3^{A_2}\gamma_4^{A_3}
, \quad \mu = 1,2,3,4,  \quad \gamma_5 = \gamma_1\gamma_2\gamma_3\gamma_4
\end{eqnarray*}
\begin{eqnarray*}
\gamma_\mu\gamma_\nu +  \gamma_\nu\gamma_\mu     = \delta_{\mu\nu} 2\mathbb{I}_2, \quad \mu,\nu = 1,2,3,4,5
\end{eqnarray*}
The unitary transformation in (\ref{2017_8_30_20}) and (\ref{2017_8_30_21}) are modified to be
\begin{eqnarray*}
 \psi^{\alpha a}(Y)  = \frac{1}{2\sqrt{2}}\sum_A \Gamma_A^{\alpha a} \chi(A,Y), \quad
 \bar \psi^{\alpha a}(Y)  = \frac{1}{2\sqrt{2}}\sum_A \bar \chi(A,Y)\Gamma_A^{*\alpha a}
\end{eqnarray*}
The kinetic part can also be written as (\ref{2017_9_25_4}) where the summation is taken for all modes
\begin{eqnarray*}
k = 2\pi \Big(\frac{m_0+\frac{1}{2}}{N_t},\frac{m_1}{N_x},\frac{m_2}{N_x},\frac{m_3}{N_x}\Big), \quad 0\leq m_0 < N_t/2, \ 0\leq m_1,m_2,m_3<N_x/2
\end{eqnarray*}
Eq. (\ref{2017_9_25_6})(\ref{2017_9_25_7})(\ref{2017_9_25_9}) are still valid except that $\mu$ runs from 1 to 3.
Eq. (\ref{2017_9_25_10})(\ref{2017_9_25_11})(\ref{2017_9_25_12}) are modified to be
\begin{eqnarray}\label{2017_9_25_15}
\sum_x D^{-1}_{x,x}  = \sum_k \frac{64m}{N(k)}
\end{eqnarray}
\begin{eqnarray}\label{2017_9_25_16}
  D^{-1}_{x^\prime,x} = \frac{1}{2} \sum_{\alpha,a,\alpha^\prime,a^\prime}\Gamma_A^{\alpha a}\Gamma_{A^\prime}^{*\alpha^\prime a^\prime}
D^{-1}_{(Y^\prime a^\prime \alpha^\prime; Y a  \alpha)}
\end{eqnarray}
\begin{eqnarray}\label{2017_9_25_17}
D^{-1}_{Y^\prime;Y} = \frac{1}{N_t/2(N_x/2)^3} \sum_k e^{i k \cdot 2(Y^\prime - Y)} D^{-1}(k)
\end{eqnarray}
respectively. The details of staggered fermion matrix can be found in the supplement material. We have checked the formula  (\ref{2017_9_13_10})(\ref{2017_9_25_11})(\ref{2017_9_25_16}) for the inverse matrices by Matlab.

 \section{The gap equation}\label{The gap equation}
The main contribution of the effective action (\ref{2016_6_23_4}) to the partition function can be obtained by the gap
equation if $N_f\rightarrow \infty$,
 \begin{eqnarray}\label{2017_3_18_2}
 \frac{\Sigma}{g^2} =  \frac{1}{N_tN_x^2} \sum_{x} D^{-1}_{x,x}
\end{eqnarray}
Here $D$ is defined in (\ref{2017_8_30_11}) where $m$ is replaced by $m+\Sigma$.  The right hand side
of (\ref{2017_3_18_2}) can be calculated from (\ref{2017_8_30_30})(\ref{2017_8_30_31}) where $m$ is replaced by $m+\Sigma$. The first derivative
of $\Sigma^2$ with respect to $\mu$ can be computed from the gap equation (For simplicity, we assume that $m=0$)
\begin{eqnarray}\label{2017_9_1_4}
 \frac{\partial\Sigma^2}{\partial \mu} =  \frac{\sum_k (\sinh 2\mu \cos 2k_0 + i \cosh 2\mu \sin 2k_0) N(k)^{-2}}{\sum_k N(k)^{-2}}
\end{eqnarray}

If the average $\Sigma$ of $\sigma$ has been calculated from the gap equation, the free energy density in the large $N_f$ limit is
$$ \ln Z = - N_tN_x^2 \frac{\Sigma^2}{2g^2} + \ln \det D  $$
where $\ln\det D = \prod_k \det D(k)$ up to a constant. The other thermodynamic quantities can be calculated. For example,
the fermion density can be analytically calculated
\begin{eqnarray}\label{2017_9_13_20}
&&  \frac{1}{N_tN_x^2} \frac{\partial\ln Z}{\partial \mu} \nonumber \\
 &=& - \frac{1}{2g^2} \frac{\partial \Sigma^2}{\partial \mu} + \frac{1}{N_tN_x^2} \Big(e^\mu \sum_x D^{-1}_{x+\hat 0,x}s^1_x + e^{-\mu} \sum_x D^{-1}_{x-\hat 0,x}s^2_x \Big) \nonumber \\
 &=& - \frac{1}{2g^2} \frac{\partial \Sigma^2}{\partial \mu} + \frac{1}{N_tN_x^2} \Big(\cosh\mu \sum_x (D^{-1}_{x+\hat 0,x}s^1_x + D^{-1}_{x-\hat 0,x}s^2_x) \nonumber \\
&& + \sinh\mu \sum_x (D^{-1}_{x+\hat 0,x}s^1_x - D^{-1}_{x-\hat 0,x}s^2_x) \Big)
\end{eqnarray}
where $\frac{\partial \Sigma^2}{\partial \mu}$, and two sums over $x$ in (\ref{2017_9_13_20}) are given in (\ref{2017_9_1_4}) (\ref{2017_9_1_4_1})  and (\ref{2017_9_1_10}), respectively. The $N(k)$ for each mode $k$ in (\ref{2017_9_1_4_1})(\ref{2017_9_1_10})(\ref{2017_9_1_4}) is given by (\ref{2017_8_30_30}) with the replacement of $m$ by $m+\Sigma$ (Here for simplicity we assume that $m=0$) and $\Sigma$ is solved from the gap equation (\ref{2017_3_18_2}).

\section{Simulation results}\label{Simulation results}

\subsection{Large volume limit}
Let us consider the large volume limit for the non-interacting theory. The partition function $Z=\int d\bar\chi d\chi  e^{-\bar\chi D\chi}  = \det D$, where the stagger fermion matrix $D$ is given by (\ref{2017_8_30_11}).  The ratio of the non-dimensional chiral condensate $a^2 \langle\bar \psi \psi\rangle$ and non-dimensional mass $m=a \tilde m$ is
\begin{eqnarray}\label{2017_12_18_2}
\frac{a^2 \langle\bar \psi \psi\rangle }{a\tilde m}  =\frac{ \langle\bar \chi \chi\rangle}{a\tilde m}
 = \frac{ \langle \sum_x \bar \chi(x) \chi(x)\rangle}{a\tilde m(N_tN_x^2)} =
 \frac{ \sum_x  D^{-1}_{x,x}}{a\tilde m(N_tN_x^2)}
 =\frac{8}{N_tN_x^2} \sum_k \frac{1}{N(k)}
 \end{eqnarray}
 where in the last equality we used Equ. (\ref{2017_8_30_31}) where $N(k)$, depending on $m$ and $\mu$, is given by (\ref{2017_8_30_30}). Note that there are $N_tN_x^2/8$ modes $k$ in (\ref{2017_12_18_2}). The ratio of the non-dimensional fermion density $a^3 \rho$ and $(a \tilde\mu)^3$
\begin{eqnarray}\label{2017_12_18_3}
\frac{a^3 \rho}{(a \tilde \mu )^3} & = &\frac{1}{\tilde \mu^3} \Big(\frac{1/\beta}{\beta L^2 }\Big) \frac{\partial \ln Z}{\partial \tilde \mu}
 =  \frac{1}{N_t\beta L^2 \tilde \mu^3} \frac{\partial \ln Z}{\partial\mu} \nonumber \\
&=& \frac{1}{N_t\beta L^2 \tilde \mu^3}\Big[ \frac{\cosh\mu}{2}8\sum_k \frac{b_1  \sin 2k_0 -  a_0 (\cos 2k_0 +1)}{N(k)} + \nonumber\\
 &  & \frac{\sinh\mu}{2} (-8 i) \sum_k \frac{a_0   \sin 2k_0 +  b_1 (\cos 2k_0-1 )}{N(k)} \Big]
 \end{eqnarray}
 where in the last equality we used (\ref{2017_9_1_4_1}) and (\ref{2017_9_1_10}).

\begin{figure}
\centering
\includegraphics[width=8cm,height=6cm]{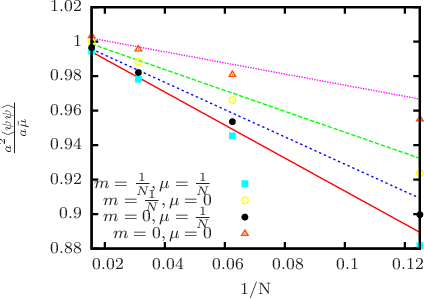}
\caption{The dependence of $\frac{a^2\langle \bar\psi\psi\rangle}{a\tilde m}$ on $N$, $N=4,8,16,32,64,128,256,512$. (1) $ m=1/N, \mu=1/N$ with fitting $-0.9563/N +  1.009$,
(2) $ m=1/N, \mu=0$ with fitting $ -0.6051/N +  1.008$,
(3) $ m=0, \mu=1/N$ with fitting $ -0.7904/N +  1.008 $,
(4) $ m=0, \mu=0$ with fitting $ -0.3224/N+1.007 $.}\label{N_limit_1}
\end{figure}

\begin{figure}
\centering
\includegraphics[width=8cm,height=6cm]{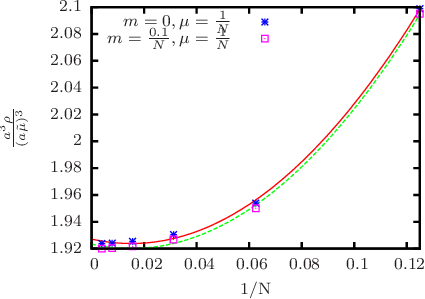}
\caption{The dependence of $\frac{a^3\rho}{(a\tilde\mu)^4}$ on $N$, $N=8,16,32,64,128,256$. (1) $ m=0, \mu=1/N$ with fitting $ 14.4370/N^2   -0.4345/N +    1.9271$,
(2) $ m=0.1/N, \mu=0$ with fitting $ 14.4288/N^2   -0.4343/N +    1.9234$.}\label{N_limit_3}
\end{figure}

     We consider the case $L=\beta$, $a=a_t$ and thus $N_x=N_t \equiv N$.  We fix $ \tilde \mu L$ and $\tilde m L$ and then calculate $\frac{a^2 \langle\bar \psi \psi\rangle }{a\tilde m}$ and $\frac{a^3 \rho}{(a \tilde \mu )^3} $ in the large $N$ limit for fixed lattice size $a$. In fact  $\frac{a^2 \langle\bar \psi \psi\rangle }{a\tilde m}$ and $\frac{a^3 \rho}{(a \tilde \mu )^3}$  does not depend on the lattice size $a$ since the non-dimensional mass $m=a\tilde m=\frac{\tilde m L}{N}$ and non-dimensional chemical potential $\mu=a\tilde \mu=\frac{\tilde \mu L}{N}$ does not depends on lattice
 size $a$. Figure \ref{N_limit_1} and shows the dependence of $\frac{a^2 \langle\bar \psi \psi\rangle }{a\tilde m}$ on $N$ with fixed $ \tilde \mu L, \tilde m L=0,1$. The linear fitting with respect to $1/N$ shows that the large $N$ limit of $\frac{a^2 \langle\bar \psi \psi\rangle }{a\tilde m}$ is close to $1.008$ for all four cases, this is because $m=1/N$ and $\mu=1/N$ both vanish for large $N$ limit.
Figure \ref{N_limit_3} shows the dependence of $\frac{a^3 \rho}{(a \tilde \mu )^3}$ on $N$, where $\tilde \mu L=1$ and $\tilde m L=0,1$. The large $N$ limit is close to $1.9271$ for $m=0$ and $1.9234$ for $m=0.1/N$, respectively.

\subsection{Phase diagram}

The phase diagram of the Gross-Neveu model in the large $N_f$ limit is well known \cite{Rosenstein_59}\cite{Rosenstein_3088}\cite{Hands_9206024}. In this limit the phase diagram of $(g^{-2}, \mu, T)$ is based on the calculation of $\Sigma$. Basically for $T=0$ and $\mu=0$, there is a critical coupling $g_c^{-2}$ such that the chiral symmetry is broken $\Sigma>0$ if the coupling is strong enough $g^{-2}<g_c^{-2}$. This critical coupling depends on the regularization of
the continuum model. For the lattice regularization in this paper, $g_c^{-2}\sim a^{-1}$ where $a$ is the lattice size. For fixed coupling $g^{-2}<g_c^{-2}$ which is not far away from the critical coupling (Otherwise, the continuum limit $a\rightarrow 0$ can not be taken),
denote $\Sigma_0$ be the value of $\Sigma$ at this coupling $g^{-2}$ with vanishing temperature $T$ and chemical potential $\mu$. The gap equation (\ref{2017_8_30_9}), which is solved exactly in the chiral limit in Ref. \cite{Hands_9206024}, shows that
there exists a critical temperature $T_c=\frac{\Sigma_0}{2\ln 2}$ such that the chiral symmetry is broken if $T<T_c$ at this coupling $g^{-2}$ and
$\mu=0$. Moreover, there is another critical chemical potential $\mu_c = \Sigma_0$ such that this symmetry is broken only if $\mu<\mu_c$ at this coupling $g^{-2}$ and $T=0$. Moreover the mean field results predict that the first order transition only exists at $T=0$ and $\mu_c$ for this coupling.

\begin{figure}
\centering
\includegraphics[width=8cm,height=6cm]{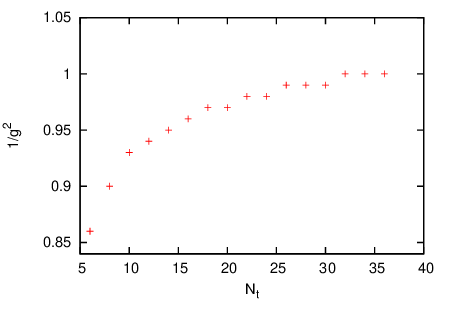}
\caption{Phase diagram of $(N_t,1/g^2)$ for $\mu=0$, $m=0$, $N_x=36$. Below the marks + is the broken phase $\Sigma>0$.}\label{gap_7}
\end{figure}

We first study the dependence of $\Sigma$ on the coupling $g$ and temperature $T=1/N_t$ without chemical potential $\mu=0$.
Figure \ref{gap_7} is the phase diagram of $(N_t,1/g^2)$ for $m=0$ and $N_x=36$. The marks + separate the symmetry phase $\Sigma=0$ (above marks) and the chiral symmetry broken phase $\Sigma>0$ (below marks). For fixed temperature $T$ there is a critical coupling $g_c^{-2}$ such that $\Sigma$ decreases to zero if $1/g^2$ is increasing to $1/g_c^2$. Figure \ref{gap_7} shows that $1/g_c^2$ is a increasing function of $N_t=1/T$ and it will close to 1 at very low temperature. On the other hand, if $g^{-2}$ is fixed, there is a critical temperature $T_c=T_c(g)$ such that $\Sigma$ is increasing from zero if $T$ is decreasing from $T_c$.

\begin{figure}

\centering
\includegraphics[width=8cm,height=6cm]{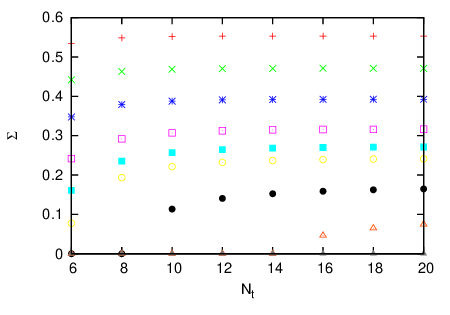}
\caption{$\Sigma$ versus $N_t$, $\mu=0$, $m=0$, $N_x=36$. $1/g^2= 0.65, 0.70, 0.75, 0.80, 0.83, 0.85, 0.90, 0.95, 1.00$ from top to bottom.}\label{gap_6}
\end{figure}

 Figure \ref{gap_6} shows the dependence of $\Sigma$ on $N_t$ for the different coupling $1/g^2$. For small $1/g^2$, for example, $1/g^2=0.65$, $\Sigma$ changes small with the temperature. For these range of parameters, it is in the deep chiral symmetry broken phase and we can not obtain the chiral symmetry phase $\Sigma=0$ even at very high temperature. For a slightly larger $1/g^2$, for example, $1/g^2=0.90$ (black dots in Figure \ref{gap_6}), we can find a transition point $T_c$, which is between $\frac{1}{8}$ and $ \frac{1}{10}$ in lattice unit. The symmetry phase and broken phase are realized for $T>T_c(g)$ and $T<T_c(g)$, respectively.

\begin{figure}
\centering
\includegraphics[width=8cm,height=6cm]{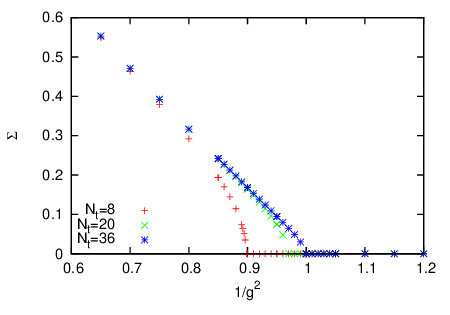}
\caption{$\Sigma$ versus $1/g^2$ for different $N_t$. $\mu=0$, $m=0$, $N_x=36$. }\label{gap_8}
\end{figure}

 Figure \ref{gap_8} shows the dependence of $\Sigma$ on $1/g^2$ at different temperature. $\Sigma$ drops continuously to 0 if $1/g^2$ is increasing to $1/g^2_c(T)$ from below, which show that the transition at the critical coupling constant $g_c(T)$ is second order. At very low temperature $T=1/N_t=1/36$, $g_c(T)$ is close to 1, which is consistent with those obtained in \cite{Hands_9208022}. This is because in the limit of $N_t,N_x\rightarrow \infty$, the gap equation at $\Sigma=0$ is reduced to
 $$ \frac{1}{g^2} = \frac{1}{N_tN_x^2} \sum_k \frac{8}{N(k)} \approx \frac{8}{\pi^3} \int_0^{\pi/2} dk \frac{1}{\sum_{\mu=0,1,2} (\cos k_\mu)^2} =1 $$

The critical temperature $T_c = \frac{\Sigma_0}{2\ln 2}$ at the coupling $g^{-2}$ and $\mu=0$ can be verified numerically. Here
we choose $N_x=36$ and $g^{-2}=0.95$ which is not too far away from the critical coupling $g_c^{-2}\approx 1$. We also choose $N_t=36$ such that it is very close to zero temperature, the value of $\Sigma$ at the zero temperature and vanishing chemical potential is $\Sigma_0= 0.0944$. To calculate
the critical temperature at this coupling, we calculate $\Sigma$ at $N_t=8,\cdots,36$ and found that $\Sigma$ is zero if $N_t$ is between 14 and 16.
 Thus the critial temperature is between $1/16=0.0625$ and $1/14=0.0667$ which is very close to  $T_c = \frac{\Sigma_0}{2\ln 2} = \frac{0.0944}{2\ln 2} = 0.0680$.

\begin{figure}
\centering
\includegraphics[width=8cm,height=6cm]{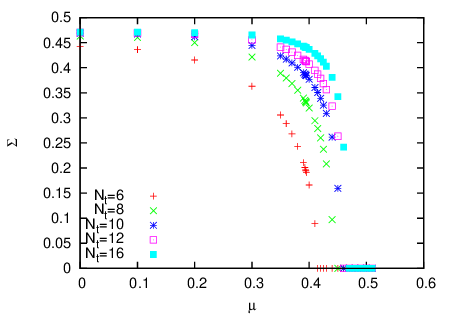}
\caption{$\Sigma$ versus $\mu$, $m=0$, $g=1.19525$ ($1/g^2=0.70$), $N_x=36$}\label{gap_2}
\end{figure}

\begin{figure}
\centering
\includegraphics[width=8cm,height=6cm]{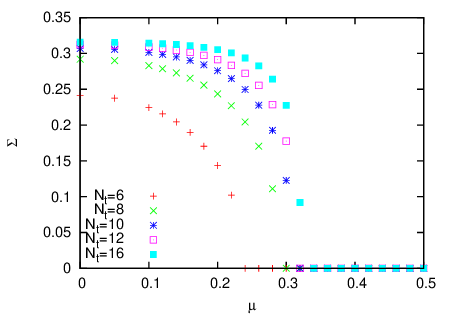}
\caption{$\Sigma$ versus $\mu$, $m=0$, $g=1.1180$ ($1/g^2=0.80$), $N_x=36$}\label{gap_3}
\end{figure}

Now let us study the effect of chemical potential on the chiral condensate $\Sigma$.  Figure \ref{gap_2} shows the dependence of $\Sigma$ on the chemical potential at the different temperature $T=1/N_t$.  $\Sigma$ drops sharply
 near $\mu_c\approx 0.45$ in the limit of zero temperature $N_t=16$, i.e., $T=1/16$, which suggest a first order transition at the zero temperature. This first order transition at the zero temperature is verified by the analytical calculation, $\mu_c = \Sigma_0$ where $\Sigma_0$ is
 the $\Sigma$ with $\mu=0$ \cite{Hands_9206024}. For the temperature $T=1/16$, $\Sigma_0\approx 0.47$ is slightly larger than $\mu_c\approx 0.45$.  If the temperature is raised, e.g., $N_t=6$, it is more difficult to find a critical chemical potential such that the chiral symmetry is restored. This is not caused by the smallness of $N_x=36$, since the our results is always obtained for $N_x=36$, which is very close to the thermodynamics limit, i.e.,
the result changes very small if $N_x$ is larger than 36. We also note that the transition at finite temperature is the second order, as explained in
 \cite{Hands_9206024}.  Figure \ref{gap_3} shows the dependence of $\Sigma$ on $\mu$ for a larger $1/g^2=0.80$. Compared with Figure \ref{gap_2}, $\Sigma$ at $\mu=0$ and the critical chemical potential in Figure \ref{gap_3} become smaller, and thus the figures in Figure \ref{gap_3} is obtained by moving those figures of Figure \ref{gap_2} in the left-down direction. For the same temperature, for example, $N_t=16$,
it is more difficult to find the critical chemical potential in Figure \ref{gap_3} than those in Figure \ref{gap_2}. Both Figure \ref{gap_2} and Figure \ref{gap_3} show that the critical chemical potential $\mu_c$ is decreased if the temperature is increased. At zero temperature, the mean field exact result show the critical chemical potential $\mu_c$ is just the value of $\Sigma_0$ at the vanishing chemical potential. This is exactly recovered in Figure \ref{gap_3} where $\mu_c = 0.32$ for $g^{-2}=0.80$ with $N_t=16$.

\begin{figure}
\centering
\includegraphics[width=8cm,height=6cm]{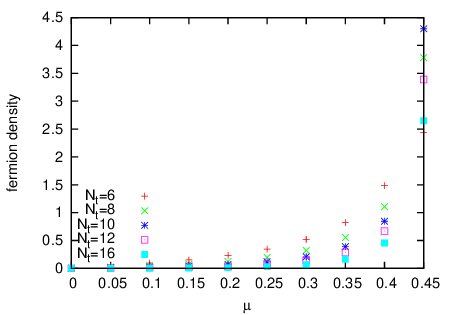}
\includegraphics[width=8cm,height=6cm]{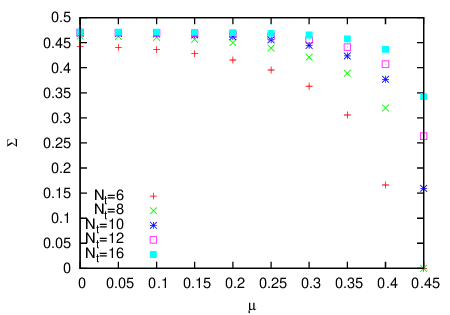}
\caption{$\Sigma$ and fermion density vs $\mu$, $m=0$, $g=1.19525$ ($1/g^2=0.70$), $N_x=36$}\label{gap_4 gap_5}
\end{figure}

Figure \ref{gap_4 gap_5} shows the dependence of $\Sigma$ and fermion density on the chemical potential at $1/g^2=0.7$. At low temperature $N_t=16$,
$\Sigma$ drops rapidly near the critical chemical potential $\mu_c \approx 0.45$, and the fermion density increase very fast, which suggest $\Sigma$ and fermion density are not continuous at $\mu_c$ at zero temperature and thus they can be regarded as the order parameters.

For finite $N_f=2,6,12$, the hybrid Monte Carlo method was used to study the symmetry breaking and symmetry restoration
\cite{Hands_9206024}\cite{Kogut_9904008}. The Monte Carlo results for $N_f=12$ is consistent with those obtained from the correction of mean field theory to  order $1/N_f^2$. The staggered fermion of Gross-Neveu model with $N_f$ flavors is equivalent to the Wilson-like fermion (4 Dirac components) with $2N_f$ flavors. To avoid the sign problem in the hybrid Monte Carlo method, $N_f$ must be even.

\begin{figure}
\centering
\includegraphics[width=8cm,height=6cm]{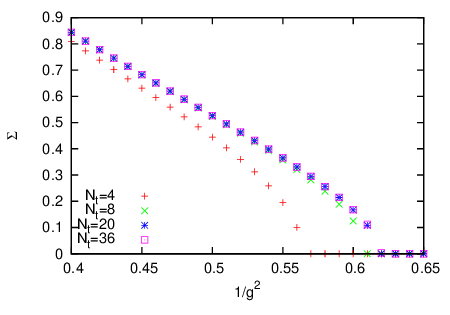}
\caption{$\Sigma$ versus $1/g^2$ for different $N_t$. $\mu=0$, $m=0$, $N_x=36$. }\label{gap_9}
\end{figure}

\begin{figure}
\centering
\includegraphics[width=8cm,height=6cm]{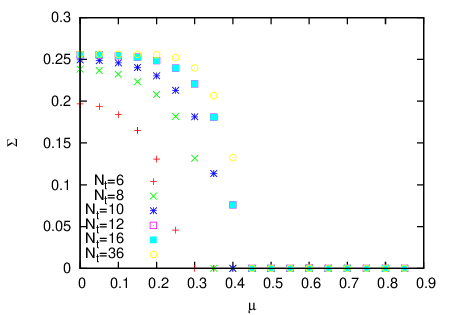}
\includegraphics[width=8cm,height=6cm]{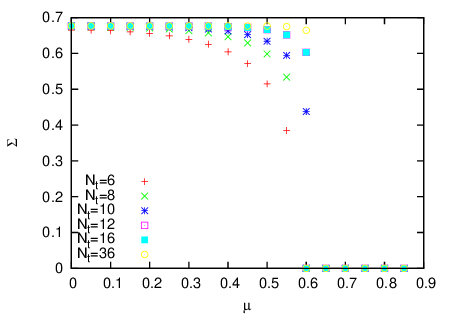}
\caption{$\Sigma$ versus $\mu$, $m=0$, ($1/g^2=0.58$), $N_x=36$. Left (3+1d), Right (2+1d)}\label{gap_10_11}
\end{figure}

For the 3+1d Gross-Neveu model, we also calculate the dependence of $\Sigma$ on the coupling and chemical potential at different temperature. Figure
\ref{gap_9} shows the value of $\Sigma$ depending on the coupling  for the vanishing chemical potential. Compared to Figure \ref{gap_8} for the 2+1d model, the critical coupling becomes smaller. Moreover, the dependence of $\Sigma$ on the temperature is less sensitive.  Figure \ref{gap_10_11} shows the dependence of $\Sigma$ on the chemical potential at the coupling $1/g^2=0.58$ for the 2+1d and 3+1d Gross-Neveu model, the critical chemical potential is smaller for the 2+1d model than those for the 3+1d model.

\section{Conclusion}\label{Conclusion}
The staggered fermion for the Gross-Neveu model at finite density and temperature are revisited. In the large $N_f$  limit, this model in 1+1d, 2+1d and 3+1d dimension can be easily solved in momentum space. Moreover an explicit formula for the inverse matrix for the 1+1d, 2+1d and 3+1d staggered fermion matrix are found, which can be implemented by parallelization. This formula can also generalized to the other space dimensions. For the odd space dimension, the orthogonal transformation were found \cite{Burden_545}. The key point to find the explicit formula for the inverse matrix is to use the properties of $\Gamma_A$ and $B_A$ as shown in Section \ref{Staggered fermion in momentum space}. These properties for the even number of space dimension is more simple, as shown in the supplement material.

The dependence of chiral condensate and fermion density on the coupling, temperature and chemical potential are obtained by solving the gap equation. Our results for the 2+1d case reproduce the analytical results. We also compared the chiral condensate for the 2+1d and 3+1d case in the same range of parameters and shows that the reason for symmetry breaking and restoration can be explained by the suitable choice of the coupling, temperature and chemical potential.

\vspace{1cm}

 Acknowledgments.
  Daming Li was supported by the
National Science Foundation of China (No. 11271258, 11971309).



{}

\appendix

\section{\label{Appendix_1}Proof of properties of $\Gamma_A$ and $B_A$}
The notations for $\{A_i\}_{i=0}^2$ in (\ref{2017_9_12_5}) is a little awkward. I replace $A_0$, $A_1$ and $A_2$ in (\ref{2017_9_12_5}) by
 $A_1$, $A_2$ and $A_3$, respectively. Thus
\begin{eqnarray}\label{2017_9_12_7}
 \Gamma_A =  \sigma_1^{A_1}\sigma_2^{A_2}\sigma_3^{A_3}, \quad
B_A =  (-\sigma_1)^{A_1}(-\sigma_2)^{A_2}(-\sigma_3)^{A_3} = (-1)^{A_1+A_2+A_3}\Gamma_A
\end{eqnarray}
The three Pauli matrices
\begin{eqnarray*}
(\sigma_1)^{\alpha\beta} = (-1)^\beta\varepsilon_{\alpha\beta}, \quad
(\sigma_2)^{\alpha\beta}=(-i)\varepsilon_{\alpha\beta} , \quad
(\sigma_3)^{\alpha\beta}=(-1)^{\beta-1}\delta_{\alpha\beta} ,\quad \alpha,\beta=1,2
\end{eqnarray*}
satisfies the completeness relation
\begin{eqnarray}\label{2017_9_12_8}
\delta_{\alpha a}\delta_{\beta b} + \sum_{\mu=1,2,3} \sigma_\mu^{ \alpha a } \sigma_\mu^{ *\beta b} = 2 \delta_{\alpha\beta}\delta_{ab}
\end{eqnarray}

We first have
\begin{eqnarray}\label{2017_9_12_37}
\sum_{A,A_3=0}\Gamma_{A}^{\alpha a}\Gamma_{A}^{*\beta b}& =& \sum_{A_1,A_2} (\sigma_1^{A_1}\sigma_2^{A_2})^{\alpha a} (\sigma_1^{*A_1}\sigma_2^{*A_2})^{\beta b}  \nonumber \\
&=& \sum_{A_1,A_2} (\sigma_1^{A_1})^{\alpha \gamma}(\sigma_2^{A_2})^{\gamma a} (\sigma_1^{*A_1})^{\beta \gamma^\prime}
(\sigma_2^{*A_2})^{\gamma^\prime b}  \nonumber \\
&=& \sum_{A_1} (\sigma_1^{A_1})^{\alpha \gamma}(\sigma_1^{*A_1})^{\beta \gamma^\prime}\sum_{A_2}(\sigma_2^{A_2})^{\gamma a}
(\sigma_2^{*A_2})^{\gamma^\prime b}  \nonumber \\
&=&  (\delta_{\alpha \gamma} \delta_{\beta \gamma^\prime} + \sigma_1^{\alpha \gamma}\sigma_1^{*\beta \gamma^\prime} )  (\delta_{\gamma a}
\delta_{\gamma^\prime b} + \sigma_2^{\gamma a}\sigma_2^{*\gamma^\prime b})  \nonumber \\
&=&  \delta_{\alpha a} \delta_{\beta b} + \sigma_2^{ \alpha a } \sigma_2^{ *\beta b}  + \sigma_1^{ \alpha a } \sigma_1^{ *\beta b} + (\sigma_1\sigma_2)^{ \alpha a } (\sigma_1^*\sigma_2^*)^{ \beta b}  \nonumber \\
&=&  \delta_{\alpha a} \delta_{\beta b} + \sigma_2^{ \alpha a } \sigma_2^{ *\beta b}  + \sigma_1^{ \alpha a } \sigma_1^{ *\beta b} + \sigma_3^{ \alpha a } \sigma_3^{* \beta b} \nonumber \\ &=& 2 \delta_{\alpha\beta}\delta_{ab} \quad \text{by } (\ref{2017_9_12_8})
\end{eqnarray}
which is also valid if $(1,2)$ is replaced by $(1,3)$ or $(2,3)$. Secondly,
 \begin{eqnarray*}
\sum_A\Gamma_{A}^{\alpha a}\Gamma_{A}^{*\beta b}& =& \sum_{A_1,A_2,A_3} (\sigma_1^{A_1}\sigma_2^{A_2}\sigma_3^{A_3})^{\alpha a} (\sigma_1^{*A_1}\sigma_2^{*A_2}\sigma_3^{*A_3})^{\beta b}  \\
&=& \sum_{A_1,A_2,A_3} (\sigma_1^{A_1}\sigma_2^{A_2})^{\alpha t}(\sigma_3^{A_3})^{ta} (\sigma_1^{*A_1}\sigma_2^{*A_2})^{\beta t^\prime}(\sigma_3^{*A_3})^{t^\prime b} \\
 &=& 2 \delta_{\alpha\beta}\delta_{tt^\prime} \Big( \delta_{ta}\delta_{t^\prime b} + (-1)^{a+b}\delta_{ta}\delta_{t^\prime b}  \Big) \\
 & = & 4 \delta_{\alpha\beta}\delta_{ab}
\end{eqnarray*}
Inserting $B_A = (-1)^{A_1+A_2+A_3}\Gamma_A$ in the above equality, we have $\sum_A B_{A}^{\alpha a}B_{A}^{*\beta b}=4 \delta_{\alpha\beta}\delta_{ab}$.

\begin{eqnarray*}
\sum_A\Gamma_{A}^{\alpha a}B_{A}^{*\beta b}& =&   \sum_{A_1,A_2,A_3} (-1)^{A_1+A_2+A_3} (\sigma_1^{A_1}\sigma_2^{A_2}\sigma_3^{A_3})^{\alpha a} (\sigma_1^{*A_1}\sigma_2^{*A_2}\sigma_3^{*A_3})^{\beta b}  \\
&=& \sum_{A_1,A_2,A_3} (-1)^{A_1+A_2}(-1)^{A_3}(\sigma_1^{A_1}\sigma_2^{A_2})^{\alpha t}(\sigma_3^{A_3})^{ta} (\sigma_1^{*A_1}\sigma_2^{*A_2})^{\beta t^\prime}(\sigma_3^{*A_3})^{t^\prime b} \\
 &=& \Big( \delta_{\alpha t} \delta_{\beta t^\prime} - \sigma_2^{ \alpha t } \sigma_2^{ *\beta t^\prime}  - \sigma_1^{ \alpha t } \sigma_1^{ *\beta t^\prime} + \sigma_3^{ \alpha t } \sigma_3^{* \beta t^\prime} \Big) \Big( \delta_{ta}\delta_{t^\prime b} - (-1)^{a+b}\delta_{ta}\delta_{t^\prime b}  \Big) \\
 & = & \Big( \delta_{\alpha a} \delta_{\beta b} - \sigma_2^{ \alpha a } \sigma_2^{ *\beta b}  - \sigma_1^{ \alpha a } \sigma_1^{ *\beta b} + \sigma_3^{ \alpha a } \sigma_3^{* \beta b} \Big)\Big( 1 - (-1)^{a+b} \Big)   = 0
\end{eqnarray*}
where in the last equality we used
\begin{eqnarray*}
&& \delta_{\alpha a} \delta_{\beta b} - \sigma_2^{ \alpha a } \sigma_2^{ *\beta b}  - \sigma_1^{ \alpha a } \sigma_1^{ *\beta b} + \sigma_3^{ \alpha a } \sigma_3^{* \beta b} \nonumber \\
&= & \delta_{\alpha a} \delta_{\beta b} - \varepsilon_{\alpha a} \varepsilon_{\beta b} -  (-1)^{a+b}
\varepsilon_{\alpha a} \varepsilon_{\beta b} + (-1)^{a+b}\delta_{\alpha a} \delta_{\beta b}   \nonumber \\
&= & \Big( 1 + (-1)^{a+b} \Big) ( \delta_{\alpha a} \delta_{\beta b}- \varepsilon_{\alpha a} \varepsilon_{\beta b} )  = 0 , \quad \text{if \ } a\neq b
\end{eqnarray*} 

  To prove that
\begin{eqnarray}\label{2017_9_12_38}
   \sum_{A,A_\mu=1}\Gamma_{A}^{\alpha a } (\sigma_\mu^*B_{A}^*)^{\alpha^\prime a^\prime} = -2(\sigma_\mu^*)^{aa^\prime}\delta_{\alpha \alpha^\prime}
\end{eqnarray}
we want to prove that
\begin{eqnarray*}
  \sum_{A,A_\mu=1}\sigma_\mu^{*ba}\Gamma_{A}^{\alpha a } (\sigma_\mu^*B_{A}^*)^{\alpha^\prime a^\prime} = -2\sigma_\mu^{*ba}(\sigma_\mu^*)^{aa^\prime}\delta_{\alpha \alpha^\prime}
\end{eqnarray*}
i.e.,
\begin{eqnarray*}
  \sum_{A,A_\mu=1}(\Gamma_{A}\sigma_\mu)^{\alpha b} (\sigma_\mu^*B_{A}^*)^{\alpha^\prime a^\prime} = -2\delta_{ba^\prime}\delta_{\alpha \alpha^\prime}
\end{eqnarray*}
This is obvious since the left hand side is
\begin{eqnarray}\label{2021_5_3_0}
 && \sum_{A,A_\mu=1}(\sigma_1^{A_1}\cdots \sigma_{\mu-1}^{A_{\mu-1}}\sigma_{\mu+1}^{A_{\mu+1}} \cdots \sigma_{3}^{A_{3}})^{\alpha b} (-1)^{A_{\mu+1}+\cdots +A_3} \nonumber  \\
 && (\sigma_1^{*A_1}\cdots \sigma_{\mu-1}^{*A_{\mu-1}}\sigma_{\mu+1}^{*A_{\mu+1}} \cdots \sigma_{3}^{*A_{3}})^{\alpha^\prime a^\prime} (-1)^{A_1+ \cdots +A_{\mu-1}}  (-1)^{A_1+A_2+A_3} \nonumber  \\
 &=& \sum_{A,A_\mu=1}(-1)^{A_\mu}(\sigma_1^{A_1}\cdots \sigma_{\mu-1}^{A_{\mu-1}}\sigma_{\mu+1}^{A_{\mu+1}} \cdots \sigma_{3}^{A_{3}})^{\alpha b}  \nonumber \\
 &&  (\sigma_1^{*A_1}\cdots \sigma_{\mu-1}^{*A_{\mu-1}}\sigma_{\mu+1}^{*A_{\mu+1}} \cdots \sigma_{3}^{*A_{3}})^{\alpha^\prime a^\prime}  \\
  & = & -2\delta_{ba^\prime}\delta_{\alpha \alpha^\prime},  \quad \text{by } (\ref{2017_9_12_37}) \ \text{if  } \mu=3 \nonumber 
\end{eqnarray}
Similarly, (\ref{2017_9_12_38}) is also valid if $A_\mu=1$ and $-2$ are replaced by  $A_\mu=0$ and $+2$, respectively. This is because 
the sign $(-1)^{A_\mu}=-1$ in (\ref{2021_5_3_0}) is replaced by $(-1)^{A_\mu}=+1$.  

Obviously,
\begin{eqnarray*}
 \Gamma_{A\pm\hat\mu} = \eta_\mu(A)\sigma_{\mu} \Gamma_A, \quad \mu = 1,2,3
\end{eqnarray*}
For example, $\mu=2$,
\begin{eqnarray*}
 \Gamma_{A\pm\hat 2} = \sigma_1^{A_1}\sigma_2^{A_2\pm 1}\sigma_3^{A_3} = \sigma_1^{A_1}\sigma_2^{A_2+ 1}\sigma_3^{A_3} = \eta_2(A)\sigma_{2} \Gamma_A, \quad \eta_2(A) = (-1)^{A_1}
\end{eqnarray*}

Finally,  we have
\begin{eqnarray*}
\frac{1}{4}\text{Tr}(\Gamma_A^\dagger \Gamma_{A^\prime} + B_A^\dagger B_{A^\prime}) = \delta_{A{A^\prime}}
\end{eqnarray*}
since the left hand side is
\begin{eqnarray*}
&&  \frac{1}{4}\text{Tr}\Big[
 \left( \begin{array}{cc}
\Gamma_A &  \\
 & B_A  \\
\end{array} \right)^{\dagger}
 \left( \begin{array}{cc}
\Gamma_{A^\prime} &  \\
 & B_{A^\prime}  \\
\end{array} \right) \Big] \\
& = &   \frac{1}{4}\text{Tr}\Big(
 (\gamma_3^{A_3}\gamma_2^{A_2}\gamma_1^{A_1})(\gamma_1^{A_1^\prime}\gamma_2^{A_2^\prime}\gamma_3^{A_3^\prime})  \Big) \\
& = &   \frac{1}{4} (-1)^{(A_1+A_1^\prime)(A_2+A_3)+(A_2+A_2^\prime)A_3} \text{Tr}\Big(
 \gamma_1^{A_1+A_1^\prime}\gamma_2^{A_2+A_2^\prime}\gamma_3^{A_3+A_3^\prime}  \Big) =\delta_{AA^\prime}
\end{eqnarray*}
where we used
$$  \gamma_\mu^i \gamma_\nu^j = (-1)^{ij}\gamma_\nu^j\gamma_\mu^i, \quad \mu\neq \nu, \ i,j=0,1,2  $$
 $$\text{Tr}(\gamma_\mu)=0, \quad \text{Tr}(\gamma_\mu\gamma_\nu)=0, \quad \mu\neq \nu, \quad
\text{Tr}(\gamma_1\gamma_2\gamma_3) = 0 $$
Here the we define $
\gamma_\mu=\left( \begin{array}{cc}
\sigma_{\mu} & 0 \\
0 & -\sigma_{\mu}  \\
\end{array} \right) (\mu = 1,2,3)$.

\section{\label{Appendix_0} The derivation of (\ref{2017_9_2_1})}
The derivation of (\ref{2017_9_2_1}) is similar to $\frac{1}{2}
\sum_{x}\eta_\mu(x)\bar\chi(x)(\chi(x+\hat\mu)- \chi(x-\hat\mu))$.
\begin{eqnarray}\label{2017_10_8_3}
& &\frac{1}{2}
\sum_{x}\bar\chi(x)(\chi(x+\hat 0)+ \chi(x-\hat 0)) \nonumber \\
& =  & \frac{1}{2}\sum_{A,A^\prime,Y}\bar \chi(A,Y)\Big( \delta_{A+\hat 0,A^\prime} (\chi(A^\prime,Y)+\chi(A^\prime,Y-\hat 0))
+  \delta_{A-\hat 0,A^\prime} (\chi(A^\prime,Y+\hat 0)+\chi(A^\prime,Y)) \Big) \nonumber \\
& =  & \frac{1}{2}\sum_{A,A^\prime,Y}\bar \chi(A,Y)\Big( \frac{\delta_{A-\hat 0,A^\prime}- \delta_{A+\hat 0,A^\prime}}{2} \partial_0\chi (A^\prime,Y))
+ \frac{\delta_{A-\hat 0,A^\prime}+ \delta_{A+\hat 0,A^\prime}}{2} \delta \chi (A^\prime,Y)  \Big) \nonumber \\
 & =  & \frac{1}{2}\sum_{A,A^\prime,Y}
  \sqrt{2}\sum_{\alpha,a}\Big(\bar u^{\alpha a}(Y)\Gamma_A^{\alpha a} + \bar d^{\alpha a}(Y)B_A^{\alpha a}\Big) \nonumber\\
 &&  \Big\{
   \frac{\delta_{A-\hat 0,A^\prime}- \delta_{A+\hat 0,A^\prime}}{2}  \sqrt{2}\sum_{\alpha^\prime,a^\prime}\Big(\Gamma_{A^\prime}^{*\alpha^\prime a^\prime} \partial_0 u^{\alpha^\prime a^\prime}(Y)  + B_{A^\prime}^{*\alpha^\prime a^\prime}  \partial_0 d^{\alpha^\prime a^\prime}(Y)\Big) +\nonumber \\
 & & \frac{\delta_{A-\hat 0,A^\prime}+ \delta_{A+\hat 0,A^\prime}}{2}  \sqrt{2}\sum_{\alpha^\prime,a^\prime}\Big(\Gamma_{A^\prime}^{*\alpha^\prime a^\prime} \delta  u^{\alpha^\prime a^\prime}(Y)  +
  B_{A^\prime}^{*\alpha^\prime a^\prime}
  \delta  d^{\alpha^\prime a^\prime}(Y)\Big)\Big\} \nonumber\\
  & = & 2\sum_{Y}\Big(\bar u^{\alpha a}(Y)(\sigma_1^*)^{aa^\prime} \partial_0 d^{\alpha a^\prime}(Y)
+\bar d^{\alpha a}(Y)(-\sigma_1^*)^{aa^\prime} \partial_0 u^{\alpha a^\prime}(Y)      +\nonumber \\ &&
\bar u^{\alpha a}(Y)(\sigma_1)^{\alpha \alpha^\prime} \delta u^{\alpha^\prime a}(Y)  +
\bar d^{\alpha a}(Y)(-\sigma_1)^{\alpha \alpha^\prime} \delta d^{\alpha^\prime a}(Y)   \Big) \nonumber\\
& = & 8\sum_{Y} \Big[ \bar q(Y)  (  i\gamma_3  \otimes \sigma_1^* )\frac{\partial_0 q(Y)}{4} +
\bar q(Y)    ( \gamma_0 \otimes \mathbb{I}_2 )\frac{\delta q(Y)}{4} \Big ] \nonumber\\
& = & 8 \sum_k \Big[ \bar q(k)  (  i\gamma_3 \otimes \sigma_1^* ) i 2^{-1}\sin (2k_0) q(k) +
\bar q(k)    ( \gamma_0 \otimes \mathbb{I}_2 )  2^{-1} [\cos(2k_0)+1] q(k) \Big ]  \nonumber
\end{eqnarray}
where
$$   \delta q(Y) = q(Y+\hat 0) + 2 q(Y) + q(Y-\hat 0) $$
In the fourth equality we used the formula like
\begin{eqnarray*}
&&\sum_{A,A^\prime}\Gamma_{A}^{\alpha a }B_{A^\prime}^{*\alpha^\prime a^\prime}
 (\delta_{A-\hat 0,A^\prime} - \delta_{A+\hat 0,A^\prime}) \\ & = & \sum_{A,A_0 = 1}\Gamma_{A}^{\alpha a }B_{A-\hat 0}^{*\alpha^\prime a^\prime}  -
\sum_{A,A_0 = 0}\Gamma_{A}^{\alpha a }B_{A+\hat 0}^{*\alpha^\prime a^\prime}  \\&=&
\sum_{A,A_0 = 1}\Gamma_{A}^{\alpha a }(-\sigma_1 B_{A})^{*\alpha^\prime a^\prime}-
\sum_{A,A_0 = 0}\Gamma_{A}^{\alpha a }(-\sigma_1 B_{A})^{*\alpha^\prime a^\prime}  \\
&=&  4 \sigma_1^{*aa^\prime}\delta_{\alpha\alpha^\prime}
\end{eqnarray*}

\section{\label{Appendix_0_1} The derivation of (\ref{2017_9_1_4_1})(\ref{2017_9_1_10})(\ref{2017_9_13_10})}
First,
\begin{eqnarray*}
&&\sum_x (D^{-1}_{x+\hat 0,x}s^1_x + D^{-1}_{x-\hat 0,x}s^2_x) \nonumber \\ & = & - \frac{\int e^{-\bar\chi D\chi} \sum_x \bar\chi (x) [\chi(x+\hat 0) + \chi(x-\hat 0) ] }{\int e^{-\bar\chi D\chi}} \nonumber  \\
& = & - \frac{\int e^{-\sum_k \bar q(k) 8 D(k) q(k)} 16 \sum_k  \bar q (k) A_+(k) q(k) }{\int  e^{-\sum_k \bar q(k) 8D(k) q(k)}} \quad \text{by } (\ref{2017_9_2_1})(\ref{2017_8_30_26})\nonumber  \\
& = & -16 \sum_k  \frac{\int e^{-  \bar q(k) 8D(k) q(k)}  \bar q(k) A_+(k) q(k) }{\int  e^{-  \bar q(k) 8D(k) q(k)}} \nonumber  \\
 & = & 16\sum_k  \text{tr} [(8 D(k))^{-1}A_+(k) ] \nonumber  \\
 & = & 2\sum_k \text{tr} [ D(k)^{-1}A_+(k)]\nonumber \\
 & = & \sum_k \frac{2}{N(k)}\text{tr} \Big\{ \Big[ m - \sum_{\mu=0,1,2} (\gamma_\mu\otimes \mathbb{I}_2) a_\mu - \sum_{c=1,2,3} (\gamma_3\otimes \sigma_{c}^*) b_c  \Big]  \nonumber \\ &  & \Big[   (  i\gamma_3 \otimes \sigma_1^* ) i 2^{-1}\sin (2k_0)  +
  ( \gamma_0 \otimes \mathbb{I}_2 )  2^{-1} [\cos(2k_0)+1]  \Big ] \Big\}  \quad \text{by } (\ref{2017_8_30_29}) \nonumber \\
   & = & \sum_k\frac{2}{N(k)} \text{tr} \Big\{ (  \mathbb{I}_4 \otimes \mathbb{I}_2) \Big(b_1  2^{-1}\sin 2k_0 -  a_0 2^{-1}(\cos 2k_0 +1)\Big)  \Big\}  \nonumber \\  & = &8\sum_k \frac{b_1  \sin 2k_0 -  a_0 (\cos 2k_0 +1)}{N(k)}
\end{eqnarray*}
Similarly,
\begin{eqnarray*}
&&\sum_x (D^{-1}_{x+\hat 0,x}s^1_x - D^{-1}_{x-\hat 0,x}s^2_x) \nonumber \\
 & = & \sum_k\frac{2}{N(k)} \text{tr} \Big\{ \Big[ m - \sum_{\mu=0,1,2} (\gamma_\mu\otimes \mathbb{I}_2) a_\mu - \sum_{c=1,2,3} (\gamma_3\otimes \sigma_{c}^*) b_c  \Big]  \nonumber \\ &  & \Big[   (\gamma_0\otimes \mathbb{I}_2)i 2^{-1}\sin (2k_0)  +
   (i\gamma_3 \otimes \sigma_{1}^*)  2^{-1} [\cos(2k_0)-1]   \Big ] \Big\} \nonumber \\
   & = & \sum_k \frac{2}{N(k)}\text{tr} \Big\{ (  \mathbb{I}_4 \otimes \mathbb{I}_2) \Big(-a_0 i 2^{-1}\sin 2k_0 -  b_1 i 2^{-1}(\cos 2k_0-1 )\Big)  \Big\}  \nonumber \\  & = & (-8 i) \sum_k \frac{a_0   \sin 2k_0 +  b_1 (\cos 2k_0-1 )}{N(k)}
\end{eqnarray*}
The inverse matrix of $D$ in (\ref{2017_8_30_11}) can be calculated as follows
\begin{eqnarray*}
&&  D^{-1}_{x^\prime,x} \nonumber \\ & = & - \frac{\int e^{-\bar\chi D\chi}  \bar\chi (x) \chi(x^\prime)}
{\int e^{-\bar\chi D\chi} } \nonumber  \\
& = & - 2 \sum_{\alpha,a,\alpha^\prime,a^\prime}\frac{\int e^{-  \bar q 8D  q }  \Big[\bar u^{\alpha a}(Y)\Gamma_A^{\alpha a} + \bar d^{\alpha a}(Y)B_A^{\alpha a}\Big]  \Big[\Gamma_{A^\prime}^{*\alpha^\prime a^\prime} u^{\alpha^\prime a^\prime}(Y^\prime) + B_{A^\prime}^{*\alpha^\prime a^\prime} d^{\alpha^\prime a^\prime}(Y^\prime)\Big]   }{\int  e^{-  \bar q 8D  q}}\nonumber  \\
& = & - 2 \sum_{\alpha,a,\alpha^\prime,a^\prime} \Big[\Gamma_A^{\alpha a}\Gamma_{A^\prime}^{*\alpha^\prime a^\prime}\frac{\int e^{-  \bar q 8D  q }   \bar q_1^{\alpha a}(Y)q_1^{\alpha^\prime a^\prime}(Y^\prime)}{\int  e^{-  \bar q 8D  q}} +\Gamma_A^{\alpha a}B_{A^\prime}^{*\alpha^\prime a^\prime}\frac{\int e^{-  \bar q 8D  q }   \bar q_1^{\alpha a}(Y)q_2^{\alpha^\prime a^\prime}(Y^\prime)}{\int  e^{-  \bar q 8D  q}} +\nonumber  \\
&  & \hspace{1.7cm} B_A^{\alpha a}\Gamma_{A^\prime}^{*\alpha^\prime a^\prime}\frac{\int e^{-  \bar q 8D  q }   \bar q_2^{\alpha a}(Y)q_1^{\alpha^\prime a^\prime}(Y^\prime)}{\int  e^{-  \bar q 8D  q}} + B_A^{\alpha a}B_{A^\prime}^{*\alpha^\prime a^\prime}\frac{\int e^{-  \bar q 8D  q }   \bar q_2^{\alpha a}(Y)q_2^{\alpha^\prime a^\prime}(Y^\prime)}{\int  e^{-  \bar q 8D  q}} \Big] \nonumber  \\
& = & \frac{1}{4} \sum_{\alpha,a,\alpha^\prime,a^\prime} \Big[\Gamma_A^{\alpha a}\Gamma_{A^\prime}^{*\alpha^\prime a^\prime}
D^{-1}_{(Y^\prime a^\prime \alpha^\prime 1; Y a  \alpha 1)} +
 \Gamma_A^{\alpha a}B_{A^\prime}^{*\alpha^\prime a^\prime} D^{-1}_{(Y^\prime a^\prime \alpha^\prime 2; Y a  \alpha 1)} + \nonumber \\
&&\hspace{1.6cm} B_A^{\alpha a}\Gamma_{A^\prime}^{*\alpha^\prime a^\prime}
D^{-1}_{(Y^\prime a^\prime \alpha^\prime 1; Y a  \alpha 2)} +  B_A^{\alpha a}B_{A^\prime}^{*\alpha^\prime a^\prime}
D^{-1}_{(Y^\prime a^\prime \alpha^\prime 2; Y a  \alpha 2)} \Big]
\end{eqnarray*}

\end{document}